\documentclass[preprint,12pt]{elsarticle}

\usepackage{graphicx}
\usepackage{amssymb}

\usepackage{lineno}

\usepackage{subcaption}
\usepackage{lscape}
\usepackage{booktabs}
\usepackage{multirow}
\usepackage{listings}
\usepackage{enumitem}
\usepackage{tikz}
\usetikzlibrary{shapes}
\usetikzlibrary{arrows}
\usetikzlibrary{decorations}
\usetikzlibrary{calc}

\tikzstyle{startstop} = [rectangle, rounded corners, minimum width=2cm, minimum height=1cm,text centered, draw=black, fill=red!30]
\tikzstyle{io} = [trapezium, trapezium left angle=60, trapezium right angle=120, minimum width=3cm, minimum height=1cm, text centered, draw=black, fill=blue!30, trapezium stretches body=true]
\tikzstyle{process} = [rectangle, minimum width=2cm, minimum height=1cm, text centered, draw=black, fill=orange!30, text width=2cm]
\tikzstyle{decision} = [diamond, minimum width=2cm, minimum height=1cm, text centered, draw=black, fill=green!30]
\tikzstyle{arrow} = [thick,->,>=stealth]

\usepackage[numbers]{natbib}
\usepackage{url}
\usepackage{aas_macros}

\journal{Astronomy and Computing}

\begin{document}

\begin{frontmatter}


\title{Efficient Source Finding for Radio Interferometric Images}



\author{Lu{\'i}s Lucas}

\address{$^1$Critical Software S.A., Portugal\\ $^2$Polytechnic Institute of Leiria, Portugal\\ $^3$University of Manchester, UK.}

\author{Tim Staley}

\address{$^1$Seyosi Ltd, UK\\ $^2$University of Manchester, UK.}

\author{Anna Scaife}

\address{University of Manchester, UK.}

\begin{abstract}
Object detection in astronomical images, generically referred to as \emph{source finding}, is often performed before the object characterisation stage in astrophysical processing work flows. In radio astronomy, source finding has historically been performed by bespoke off-line systems; however, modern data acquisition systems as well as those proposed for upcoming observatories such as the Square Kilometre Array (SKA), will make this approach unfeasible. One area where a change of approach is particularly necessary is in the design of fast imaging systems for transient studies. This paper presents a number of advances in accelerating and automating the source finding in such systems. 
\end{abstract}


\end{frontmatter}


\section{Introduction}
\label{sec:intro}

Object detection, known as \textit{source finding} in radio astronomy, is a component of the standard analysis pipeline for most astronomical surveys. Source finding is used to identify individual astrophysical objects in survey data, while object characterisation estimates their properties such as location, intensity and shape. Once extracted, these properties are then summarised to form radio survey catalogues. These catalogues are constructed with the intention of providing a compressed version of the information contained in the survey and hence of the astrophysical systems represented in those data. 

The software packages used for this process are known as source finders and typically provide object detection and characterisation capabilities to produce the catalogues of detected objects as their output. 
Historically, due to the faintness of the radio sky and the comparative sparsity of extragalactic objects in radio surveys, the development of these source finders has focused largely on statistical completeness (contain all sources present in the image) and reliability (all sources found and extracted are real; \cite{hancock_aegean}), rather than optimization of the algorithms for run time and memory usage. Consequently, the source finding algorithms developed for radio astronomy over the last few decades have been designed to be highly reliable, returning lists of sources that are expected to be complete and include only a small or zero fraction of false positives. 

For the new generation of radio telescopes such as MeerKAT \cite{MeerKAT}, ASKAP \cite{ASKAP} and ultimately the Square Kilometre Array (SKA; \cite{SKA}) the improvements in the sensitivity of radio telescopes, coupled with increased survey speeds due to larger instantaneous fields-of-view (FOV), radio astronomy is starting to produce large area, sensitive maps of the sky at high cadence. As this trend increases, the data processing for these facilities will need to be fully automated, with minimal manual input, and be able to operate reliably and quickly on large volumes of data. Consequently, source finding algorithms will not only need to be accurate but will also be required to work efficiently on much larger image datasets as a matter of routine.  

One specific case where this will be required is for the slow transients pipeline (STP) of the SKA telescope. This imaging pipeline is designed to identify transient astronomical sources, the brightness of which varies on timescales long enough for them to be detected in individual residual snap-shot images from the telescope, where a residual snapshot image is defined as an image that can be created from $\sim 1$\,second of data subtracted by the sky model data. The STP is expected to produce images at this cadence and identify any transient sources in those residual images. However, the FOV (in terms of number of pixels) for the SKA will make it impossible to store each individual residual snapshot image and consequently the processing for such a pipeline must be performed in soft real time. 

When searching for such short-lived radio transients it is not possible to re-observe in order to recover objects missed by initial processing, hence accurate source finding is a key component of this processing. Consequently, to perform the processing for the SKA STP requires not only the imaging of the data to be performed efficiently, but also the source finding in each residual snap-shot image to be optimized for speed. 

In this paper we describe a new algorithm for computationally efficient source finding in interferometric radio images: {\sc ceres}. This algorithm has been designed as part of the development for the SKA slow transients pipeline with its codebase being available on GitHub \cite{STP_website}. The source finding is a component of the STP workflow being applied to the residual dirty image generated previously in the imager stage specifically by computing the residual visibility data (\emph{i.e.} by subtracting the known sky model visibility data), gridding the residual visibility data and computing the FFT of the gridded data. In the SKA STP, this procedure should be performed at $\sim 1$ second cadence.

An important feature of the algorithm developed is the ability to find both positive and negative sources in the residual dirty image. 
The potential to search for negative sources is important for imaging applications in radio astronomy that monitor fading transient sources, as well as those that use polarization data, since polarized sources can appear as negative features in Stokes images.
While the algorithm discussed performs both object detection and characterisation, it is important to recall that this paper is focused on the detection stage, for which significant speedup techniques are proposed.

The paper proceeds as follows: in \S~\ref{sec:context} we give an overview of current source finding algorithms and the computational steps required in the source finding process, in \S~\ref{sec:fast} we describe the changes and optimizations in each step that have been implemented in this new algorithm, and in \S~\ref{sec:tests_benchmarks} we present benchmarking of the {\sc ceres} algorithm in comparison with other widely used source finders. In \S~\ref{sec:conclusions} we draw our conclusions.

\section{Source Finding in Radio Interferometry}
\label{sec:context}

\subsection{Image-plane vs uv-plane source analysis }
Source finding is the process of analysing data to separate signal from noise and infer the flux intensity, position, and (in the case of extended sources) structure of radiation-emitting sources on-sky. For optical data this analysis is performed on pixelated images, and essentially consists of locating clusters of pixels which are outside the expected range of values recorded due to noise alone, then processing each cluster to estimate the required source parameters. Knowledge of the system characteristics may be used to enhance the analysis (e.g. noise-suppression with matched-filters, deconvolution using knowledge of the point-spread function to better parameterise adjacent sources, etc).

Source finding in radio-synthesis maps has evolved in parallel to optical-data processing techniques, but has a distinct set of challenges and requirements. Firstly, the transform of radio-data from recorded antenna-voltages through correlation and Fourier-transform to the image-map produces images with a strong inter-pixel noise correlation on the scale of the synthesized beam. Second, incomplete coverage of the \textit{uv}-plane may produce a synthesized beam with a complex, large-scale, multimodal structure, where the secondary peaks are known as `side-lobes'. As a result, a bright source may obscure or distort the brightness-profile of a a fainter one even if their sky-positions are separated by many multiples of the synthesized-beam width.

The latter problem is often addressed through deconvolution, typically using a variant of the {\sc clean} algorithm \cite{CLEAN}. This ameliorates the issue of side-lobes, at the cost of adding an additional, iterative, non-linear processing step to the data, which may introduce additional degeneracies and can result in significantly different images depending on the clean-parameters. Source finding algorithms are then typically run against the cleaned image.

In some circumstances - particularly if the uv-plane coverage is very sparse, or if the source has excellent signal-to-noise ratio and structure of interest on small scales, it may be more effective to model the sources at the level of the visibility data directly. In \cite{martividal_uvmultifit}, the authors  discuss a procedure to `degrid' a sky-model to the visibility plane in order fit model parameters through least-squares residuals in the visibilities. However, this requires a prior `guess' at a suitable model of the source structure, and only provides best fit parameters with limited information about the suitability of the model. \cite{lochner_biro} takes the concept further and use a Monte-Carlo process to perform a full Bayesian inference on both the the source parameters and the systematic noise effects, but are still limited to simple models (single point-source, two point-sources, or an extended Gaussian). Consequently traditional sourcefinding is still the only option for most blind-field surveys, and we limit the rest of this discussion to standard image-plane techniques.

\subsection{Image-plane source finding algorithms}
\label{sec:other_sourcefinders}
There are several widely-used source finding software packages in astronomy which take a largely similar approach. Here we give an overview of the steps involved, and briefly cover differences between the following selection: Bertin's source-extractor (`S-Extractor') \cite{bertin_sextractor}, the LOFAR-TKP Python Source-Extractor (`PySE') \cite{spreeuw_pyse, swinbank_trap}, and Hancock's `Aegean' package \cite{hancock_aegean_2018}.

\subsubsection{Background and noise estimation}
The first step in the source finding algorithms is to characterise the background and noise levels of an image, in order to choose a sensible threshold for determining pixels of interest. This step can be remarkably compute intensive due to the need to process all the image-pixels and ideally discard pixels containing significant source-flux, as discussed in \S\ref{sec:fast}. 
Generally, we expect the background pixel distribution to be well approximated by a normal distribution, so this is equivalent to estimating the mode and root mean square (RMS) of the background-pixel sample. Usually some form of sigma-clipping is applied to reject outlier pixels containing source flux. The RMS and mode can then be estimated from the remaining pixels, with either the mean, median, or some combination of both being used to estimate the mode \citep[see e.g.][]{bertin_sextractor}. For radio-synthesis maps the situation is complicated by the strong inter-pixel noise correlation, and so the routine may be modified to only use a sparse sampling of the pixels for RMS-estimation, as in Aegean, or to apply a correction factor based on knowledge of the synthesized-beam parameters, as in PySE \cite[see][\S4.3.1 therein]{swinbank_trap}.

\subsubsection{Connected-component labeling}
\label{sec:ccl}

Once the background and noise-variance are known we can pick a suitable threshold-level to reject most background-noise pixels. Typical values are a factor 5--10 times the RMS above the background level (referred to as a 5--10$\sigma$ threshold in the following text). Alternatively a threshold may be chosen by specifying a desired false-detection rate \cite{benjamini_fdr, hopkins_fdr}. Using the threshold of choice, the pixels can be converted to a binary-map (above/below threshold), which is then subdivided into labelled regions representing sources or clusters of sources. The above-threshold regions are often referred to as `islands', by way of analogy to islands of land sticking out of the ocean. This binary-map labelling procedure is more generally known as `connected components labelling', and has a fairly extensive treatment in the computer-science literature. S-Extractor uses a custom implementation of Lutz's algorithm \cite{lutz_onepass_labelling}, while PySE and Aegean make use of the SciPy function \texttt{scipy.ndimage.measurements.label} \cite{scipy}. One issue with this approach as described --- using a single threshold --- is that an island may quite frequently consist of a single pixel above the threshold level, for a source close to the detection limit. This makes it difficult to estimate any parameters other than the value and location of the island's peak pixel. PySE addresses this edge-case by making use of two thresholds referred to as the `detection and analysis'. The `detection' threshold is chosen at the same 5--10$\sigma$ level as before, while the `analysis' threshold is chosen at a lower level that would result in many more false detections if used alone, say 3--4$\sigma$. The thresholding and connected-component labelling procedure is performed using the \textit{analysis} threshold, but any islands with a peak value less than the \textit{detection} threshold are rejected as marginal detections likely due to noise fluctuations.

\subsubsection{Deblending}
One failing of the `island-finding' approach to source-extraction is that multiple bright sources located in close proximity may result in an above-threshold connected region that encompasses several sources. To circumvent this issue, all three of the sourcefinders from the literature implement some form of user-configurable `deblending' step, wherein every island is (optionally) further analyzed to determine if it can be plausibly broken up into multiple `sub-islands'. S-Extractor and PySE do this by allocating a number of thresholds on a logarithmic spacing between the original threshold and the island peak, and then re-running a connectivity analysis on the island using each of the higher thresholds in turn. If multiple sub-islands are detected at this new, higher threshold then further criteria are applied (contained flux as a portion of the original island-flux and peak-value in the sub-island) before decided whether to accept the proposed island-split. The process then repeats at the next (higher) threshold, sub-dividing further as appropriate, for a user-defined number of iterations \cite{bertin_sextractor},\cite[][pp 40--41]{spreeuw_pyse}. Aegean takes a novel alternative approach, computing a map of local curvature across the image and the designating connected-regions of negative curvature (convex areas) as `summits'. These `summits' are then used in conjunction with the standard threshold-map to designate sub-island regions for separate Gaussian component fitting \cite[][\S7 therein]{hancock_aegean}.

Since the work described in this paper is primarily focused on high-speed transient-detection in residual images we do not expect multiple close sources, and so have not implemented a deblending step, but the code could be extended in a straightforward manner to apply either deblending method if required for regular blind-field surveys.

\subsubsection{Island characterization and fitting}
Once island regions have been identified we can attempt to extract summary metrics about each island (hopefully corresponding to a single source). At the simplest level, we can use the position and value of the brightest pixel within an island, but this is a rather crude characterisation. The approach adopted by by S-Extractor, and refined in PySE, is to use the first and second statistical moments of the island pixels, which in the case of a Gaussian profile correspond to the mean and variance of the Gaussian (we can also estimate the rotation angle through some reparameterisation formulae). It is worth noting that sources with flux-level close to the cut-off threshold (and therefore with small islands of designated pixels) will result in underestimates of the variance (similar to a truncated Gaussian); \cite{spreeuw_pyse} derives correction factors for this systematic bias and these correction factors are applied in PySE. S-Extractor then returns these moment-estimate parameters and halts. However, it is often possible to improve accuracy through least-squares fitting of a Gaussian profile, particularly for compact sources at high signal-to-noise ratio \cite[see e.g.][figures 2.4--2.5]{spreeuw_pyse}. PySE implements Gaussian fitting as a secondary step, using the moments-parameter estimates as initial parameters for the fit and applying \texttt{scipy.optimize.leastsq}, a wrapper around the MINPACK implementation of \texttt{lmdif} (a Levenberg-Marquardt optimisation algorithm). Aegean also implements Gaussian fitting, employing some sensible constraints on the Gaussian parameter values, while simply setting the initial parameters using the peak pixel position and flux, with initial beam parameters set to the synthesized beam profile. Aegean makes use of another Python library, `lmfit', which extends the \texttt{scipy.optimize.leastsq} routine to provide fitting under parameter constraints.

\section{Efficient Source Finding}
\label{sec:fast}

Source finding is the final stage of an imaging workflow. For sparse residual images, such as those proposed for the SKA slow transients pipeline (STP), or data with high filling factors in the \emph{uv}-plane, this detection step may be performed directly over the dirty image; the only direct data dependency between the previous work flow elements and the source finding stage.
  
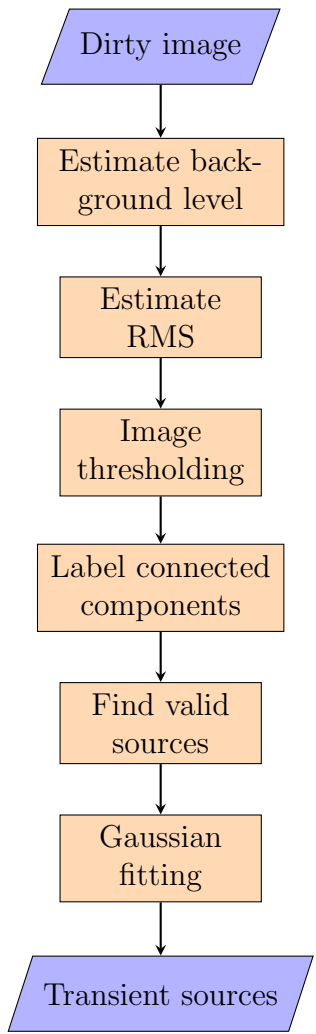
\begin{figure}[!t]

\centering
\begin{tikzpicture}[node distance=1.8cm]

\node(in1)[io] {Dirty image};
\node (pro1) [process, below of=in1, text width=3cm] {Estimate background level};
\node (pro2) [process, below of=pro1, text width=2.4cm] {Estimate RMS};
\node (pro3) [process, below of=pro2, text width=2.4cm] {Image thresholding};
\node (pro4) [process, below of=pro3, text width=3cm] {Label connected components};
\node (pro5) [process, below of=pro4, text width=2.4cm] {Find valid sources};
\node (pro6) [process, below of=pro5, text width=2.4cm] {Gaussian fitting};
\node(out1)[io, below of=pro6] {Transient sources};
\draw [arrow] (in1) -- (pro1);
\draw [arrow] (pro1) -- (pro2);
\draw [arrow] (pro2) -- (pro3);
\draw [arrow] (pro3) -- (pro4);
\draw [arrow] (pro4) -- (pro5);
\draw [arrow] (pro5) -- (pro6);
\draw [arrow] (pro6) -- (out1);

\end{tikzpicture}

\caption{Block diagram for source finding.}
\label{fig:sourcefind_diagram}

\end{figure}

The relevant steps of the source finding algorithm addressed in this work are represented in the block diagram of Figure \ref{fig:sourcefind_diagram}. The algorithm as presented in this paper can be used to find both positive and negative sources in the residual dirty image, although the search for negative sources can be disabled. Searching for negative sources is important in STP since the polarized sources can appear as negative features in Stokes images.

The main bottleneck of a source finding algorithm is generally considered to be the RMS estimation, mainly due to the sigma clipping step performed internally, which requires several passes over the image data. However, other components of source finding, such as the background level estimation and the connected component labelling also present a noticeable computational complexity which can be ignored. 

To minimise the run time, we here propose a number of optimisations and improvements to the standard source finding process. Specifically we focus on reducing the number of passes over image data, avoiding image copies and redundant computations, and taking advantage of parallel processing. 

In the following sections, we describe each step of the source finding algorithm described pictorially in Fig.~\ref{fig:sourcefind_diagram} and we present the proposed solutions for an efficient implementation using the C++ language. 

\subsection{Background level estimation}
\label{sec:backest}

The first step in source finding is to estimate the background level of the image by computing the median of the image.
The fastest existing algorithm to compute the median is \emph{quickselect} \cite{Floyd_quickselect}. It consists of a selection algorithm that finds the $k^{th}$ smallest element from an array of $n$ elements. When $n$ is odd the median is the $k^{th}$ smallest element with $k=(n+1)/2$, and when $n$ is even it is the mean of the elements $k=n/2$ and $k=n/2+1$.
The \emph{quickselect} algorithm is derived from the well known sorting algorithm \emph{quicksort}. In brief, \emph{quickselect} chooses a \emph{pivot} element $p$ from the array and rearranges the array so that all elements smaller than $p$ are to its left and all elements larger than $p$ are to its right. It then acts recursively on one of the sub-arrays (left or right), depending on where the median element lies.

The \emph{quickselect} algorithm can perform efficiently in practice and when the \emph{pivot} element is selected randomly from the array, it presents $\mathcal{O}(n)$ average computational complexity. The main disadvantage of \emph{quickselect} is the use of in-place partitioning, which rearranges the input array. In applications like imaging, where the input array is required to keep its original order, it is necessary to make a scratch copy of the array. 

Typically, in C++ programs, the median function is implemented using the \emph{Nth\_element} function provided in C++ standard library for $k^{th}$ smallest element selection based on \emph{quickselect} algorithm. The parallel implementation of the C++ standard library \cite{gnu_std_parallel} includes a parallel version of the \emph{Nth\_element} function that can be used for faster median computation.

An alternative approach for median computation is based on the work of \cite{Tibshirani_binmedian}, where two methods for fast median computation are described based on the concept of successive binning. The first method, referred to as \emph{binmedian}, presents $\mathcal{O}(n)$ average computational complexity, similarly to \emph{quickselect}. Its main advantage relative to \emph{quickselect} is the significantly faster update of the median as more data are added to the array. However, in the context of source finding application, this is feature not relevant.
The second method, referred to as \emph{binapprox}, computes an approximate median value. It tends to perform faster than \emph{binmedian}, presenting a worst-case complexity of $\mathcal{O}(n)$.

In this work, both the \emph{binapprox} method and a simplified version of \emph{binmedian} were implemented and evaluated.
The principle behind these methods is based on the following lemma (proof in \cite{Tibshirani_binmedian}):

\vskip .1in
\noindent
\newtheorem{theorem}{Theorem}
\newtheorem{lemma}[theorem]{Lemma}
\begin{lemma}
\noindent
If X is a random variable having mean $\mu$, variance $\sigma^2$, and median $m$, then $m \in [ \mu-\sigma,\mu+\sigma]$
\end{lemma}

\vskip .1in
In this context, the basic algorithm of \emph{binapprox} is given by:
\begin{enumerate}
 \item Compute the mean $\mu$ and standard deviation $\sigma$;
 \item Form $B$ bins across $[\mu-\sigma,\mu+\sigma]$ and map each data point to a bin;
 \item Find the bin $b$ that contains the median. This consist in summing the size of each bin until the result goes by the middle value, \emph{i.e.} the total number of samples divided by 2;
 \item Return the midpoint of bin $b$.
\end{enumerate}

The estimated median can differ from the exact median value by at most half the width of the interval, or $\sigma/B$. Typically, the number of bins is $B=1000$, which results in a maximum error of $1/1000^{th}$ of a standard deviation. The main advantage of \emph{binapprox} is not only its reduced running times, but also the fact that the input data are not modified, contrary to \emph{quickselect}. Conversely, its main disadvantage is that the computed median may not correspond to the true median. However, given that the maximum error is $1/1000^{th}$ of the image noise it should be a quite small and irrelevant error.

The first three steps of the \emph{binmedian} method are similar to those of \emph{binapprox}; however, instead of returning the midpoint of bin $b$ in the fourth step, \emph{binmedian} recurses on the set of samples mapped to $b$. This recursion is applied until the number of remaining samples in median bin $b$ is smaller than a fixed constant. The \emph{quickselect} method is then used to find the median based only on the few remaining samples. More details of this approach are given in \cite{Tibshirani_binmedian}.

Although it presents $\mathcal{O}(n)$ average computational complexity, the \emph{binmedian} method requires additional memory access and copy operations that reduce its performance. At each iteration, the full image array needs to be accessed to perform binning of the selected bin samples spread along the array. To avoid multiple accesses to the full array, the samples of the selected bin can be copied into a smaller array. However, this solution implies data copies at each iteration of the algorithm. 

In this work it was observed that stopping successive binning after the first iteration and using \emph{quickselect} over the remaining samples of the selected bin performed more efficiently most of the times. 
This solution can be viewed as an extension to the \emph{binapprox} method. Instead of returning the midpoint value of the selected bin, all the samples of the selected bin are copied to a new array and inputted to the \emph{quickselect} function that returns the median value. Hereafter, we will use the term \emph{binmedian} to refer to this modified solution for computing exact median.
The main advantage of this method, relative to the use of \emph{quickselect} on the original array is that the binning step significantly reduces the number of samples to be handled by \emph{quickselect}.

A common characteristic of the \emph{binapprox} and \emph{binmedian} methods is the calculation of the mean and standard deviation quantities in the first step of their algorithms. Since the standard deviation is also required for the sigma clipping step performed in the context of RMS estimation, this quantity can be reused thus avoiding redundant computations in the source finding algorithm. Consequently, in addition to the reduced running time for median computation when compared with \emph{quickselect}, the \emph{binapprox} and \emph{binmedian} functions proposed here allow for a reduction of the computational complexity in the RMS estimation step, by reusing the precomputed standard deviation quantity.

A further advantage of the binning methods is that they can be easily implemented using multi-core processing. We used the Intel Threading Building Blocks (TBB) \cite{Reinders_TBB} library to perform the binning step in parallel, specifically by assigning distinct threads to separated regions of the image and then combining the per-thread binning results. A parallel approach was also used to estimate the mean and standard deviation of the image specifically by computing the auxiliary quantities (the accumulation values) for distinct regions of the image in parallel and then combining the results, as discussed in \S~\ref{sec:rms}. To minimise the number of passes over the image, the mean and standard deviation are calculated jointly using a single reading of the image.

For source finding execution, the median function can be selected from the input configuration file. While the \emph{binapprox} method does not compute exact median, having a negligible error, the alternative \emph{binmedian} function provides the exact median at the cost of a higher computational complexity.

\subsection{RMS estimation}
\label{sec:rms}

Unless an RMS value is provided a priori as input to the source finding function, estimation of the RMS is performed globally as part of the source finding process. For RMS estimation, sigma clipping is first applied to remove the outlier samples. The sigma clipping function requires computing the standard deviation statistical quantity and subtracting the background level (median) from the image. Since the background level was previously estimated, its value is directly used by the sigma clipping function. Furthermore, as described in \S~\ref{sec:backest}, the binning-based approach used for background level estimation also computed the standard deviation quantity in its first step. This value is thus kept in memory and passed to the RMS estimation function for the sigma clipping function.

Sigma clipping is the main bottleneck of the RMS estimation procedure. This function iterates over all data, rejecting samples that are discrepant by more than a specified number of standard deviations from a central value, the median. By default, we use 5 iterations for sigma clipping, each resulting in one pass through the data. At each iteration the standard deviation is recomputed, after rejecting the discrepant samples from the calculations. 

Here we propose an efficient solution for sigma clipping and RMS estimation that reduces both the number of computations and memory writes. 
Since the input data image should not be modified, the clipped samples at each iteration of sigma clipping cannot be removed or marked in the image, for instance by using the NaN (Not-a-Number) symbol. A copy of the image could be used for this purpose but, due to its large size and the required number of writes, this solution is not recommended. Instead, the proposed solution compares all image samples, including discrepant samples detected in previous iterations, against the current and previous threshold values for sigma clipping. 
Although this approach requires additional comparisons in order to know which samples were clipped at previous iterations, it is less costly than creating a scratch copy of the image and removing or marking the clipped samples at each iteration.

Given that sigma clipping recomputes the standard deviation at each iteration using data that are quite similar, with only slight differences due to the clipped samples, we here consider a computationally efficient solution that minimises the number of computations. This proposed method uses auxiliary accumulation variables that are updated each time a sample is clipped.

For an array of $n$ samples, the standard deviation ($\sigma$) can be expressed as:

\begin{equation}
 \sigma^2 = \Bigg( \frac{1}{n} \underbrace{\sum_{i=1}^n x_i^2}_{\rm Acc2} \Bigg) - \Bigg( \frac{1}{n} \underbrace{\sum_{i=1}^n x_i }_{\rm Acc1} \Bigg)^2
\end{equation}

As can be seen from this equation, the standard deviation is computed using two auxiliary accumulation values, specifically \emph{Acc1} and \emph{Acc2}. Considering this fact, we may subtract the samples clipped at each iteration from the accumulation variables to recompute the standard deviation quantity. This solution minimises the number of computations required to update the standard deviation quantity, since it does not require one to sum all the point values and their squares repeatedly at each iteration. Furthermore, the computation of these auxiliary accumulation values can be easily performed in parallel, specifically by having multiple threads summing separate regions of the image into local auxiliary accumulation values that are combined in the end.
For the first iteration of sigma clipping, these accumulation variables are derived from the mean and standard deviation quantities previously computed during background level estimation.

As an additional optimisation, the sigma clipping and RMS estimation functions are merged. Since the RMS corresponds to the standard deviation of the clipped data, we use the standard deviation value computed in the last iteration of sigma clipping function as the estimated RMS value. 

A disadvantage of this method is that it relies on a standard deviation equation that is not considered numerically stable. This equation can suffer from loss of precision due to the difference in magnitude between one sample and the sum of all samples in the accumulation variables.
A numerically stable equation for the standard deviation is:
\begin{equation}
 S_n = S_{n-1} + (x_n-\mu_{n-1})(x_n-\mu_n),
\end{equation}
assuming
\begin{equation}
 S_n = n\sigma_n^2.
\end{equation}
and with $\mu$ being the mean quantity that can be computed using the following numerically stable equation:
\begin{equation}
 \mu_n = \mu_{n-1}+\frac{1}{n}(x_n-\mu_{n-1}),
\end{equation}

However the drawback of using these equations is their high impact on the computational performance. It can be observed that these expressions involve a larger number of operations, in particular multiplication and division operations, that would result in a longer running time. In the interest of maintaining the performance we do not currently implement this more stable method.

In summary, we use an efficient implementation of the RMS estimation function, which only requires 5 passes through the image, one per iteration of the sigma clipping function. For a faster performance, we use multi-threaded processing in which different threads compute the auxiliary accumulation values for separated regions of the image and then combine the results.
The number of performed computations is minimised by making use of the previously computed standard deviation and median quantities determined for background level estimation and adopting an efficient solution to update their values when some samples are clipped. Furthermore, the algorithm is designed so that the input image matrix does not need to be copied.

\subsection{Source Labeling}
\label{sssec:labelling}

Based on the estimated RMS value, the analysis and detection threshold values are determined. 
The image is then thresholded using the analysis threshold and the connected components (the sources) are labeled. Finally, the detection threshold is used to identify valid sources, following the same method as described for PySE in \S~\ref{sec:ccl}.

A Connected Component Labeling \cite{Samet_CCL_1988} algorithm is used to search all the regions (connected components) in the binary image generated by the thresholding step, assigning a unique label to each identified region. A connected component in a binary image is an area of pixels with value 1, which are connected per a pre-defined connectivity type. In astronomical source finding, the 8-connectivity type is used, i.e. it is assumed that two samples are connected if they are adjacent in the horizontal, vertical or diagonal directions.
There are several CCL algorithms proposed in literature, some of which exploit parallel processing \cite{Jeong_ISCIT2010,Cabaret2016}. 
The existing algorithms can be classified into three major groups:
\begin{description}
 \item [Multi-pass] algorithms usually require several passes over the image before reaching the final labels. The number of passes depends on the content. These algorithms may use specific techniques to reduce the number of passes, such as a label connection table.
 \item [Two-pass] algorithms require only two passes over the image. These algorithms usually operate in three distinct phases:
 \begin{itemize}
  \item Scanning phase: the image is scanned in the first pass to assign provisional labels to object pixels and to record the equivalence information among provisional labels.
  \item Analysis phase: this phase analyses the label equivalence information to determine the final labels.
  \item Labelling phase: this phase assigns final labels to object pixels using a second pass through the image.
 \end{itemize}
 \item [One-pass] algorithms scan the image to find an unlabelled object pixel and then assign the same label to all connected object pixels. As they perform a single pass over the image, these methods require irregular access to pixels, which usually leads to poor performance.
\end{description}

The computational performance of CCL algorithms depends not only on their approach and the number of passes over the image, but also on the input data characteristics (e.g. the number of connected components present in the image). The YACCLAB (Yet Another Connected Components Labeling Benchmark; \cite{yacclab_website}) is one example of a C++ open source framework which evaluates various CCL algorithms for different data types. 

\subsubsection{Efficient Component Labeling}
\label{sec:effccl}

Here we implement a multithreaded version of the 8-connectivity CCL algorithm provided in OpenCV \cite{opencv_library}, a reference open source library for computer vision. This algorithm is based on the approach of \cite{Wu_CCL_openCV}, which is the two-pass CCL algorithm briefly described in \S~\ref{sssec:labelling}. Although this solution requires two passes over the image, it tends to be significantly faster than one-pass algorithms that use irregular and slow memory access patterns. 
In addition to the use of multithreaded processing for CCL, the overall source find algorithm performance was improved by merging some related source finding steps into the processing loop of the two-pass CCL function. These optimisations are summarised here and described in more detail below:
\begin{enumerate}
 \item Label non-shifted FFT output image as if its quadrants were properly shifted;
 \item Use multi-threaded implementation for faster performance;
 \item Merge the thresholding step into the labelling function to avoid allocating memory for the binary image;
 \item Merge the valid source detection step with the last stage of the labelling function to save one full pass over the label map;
 \item Merge labelling of positive and negative sources in a single CCL procedure using only two image passes;
\end{enumerate}

\begin{figure*}
\centering
\includegraphics[width=0.9\textwidth]{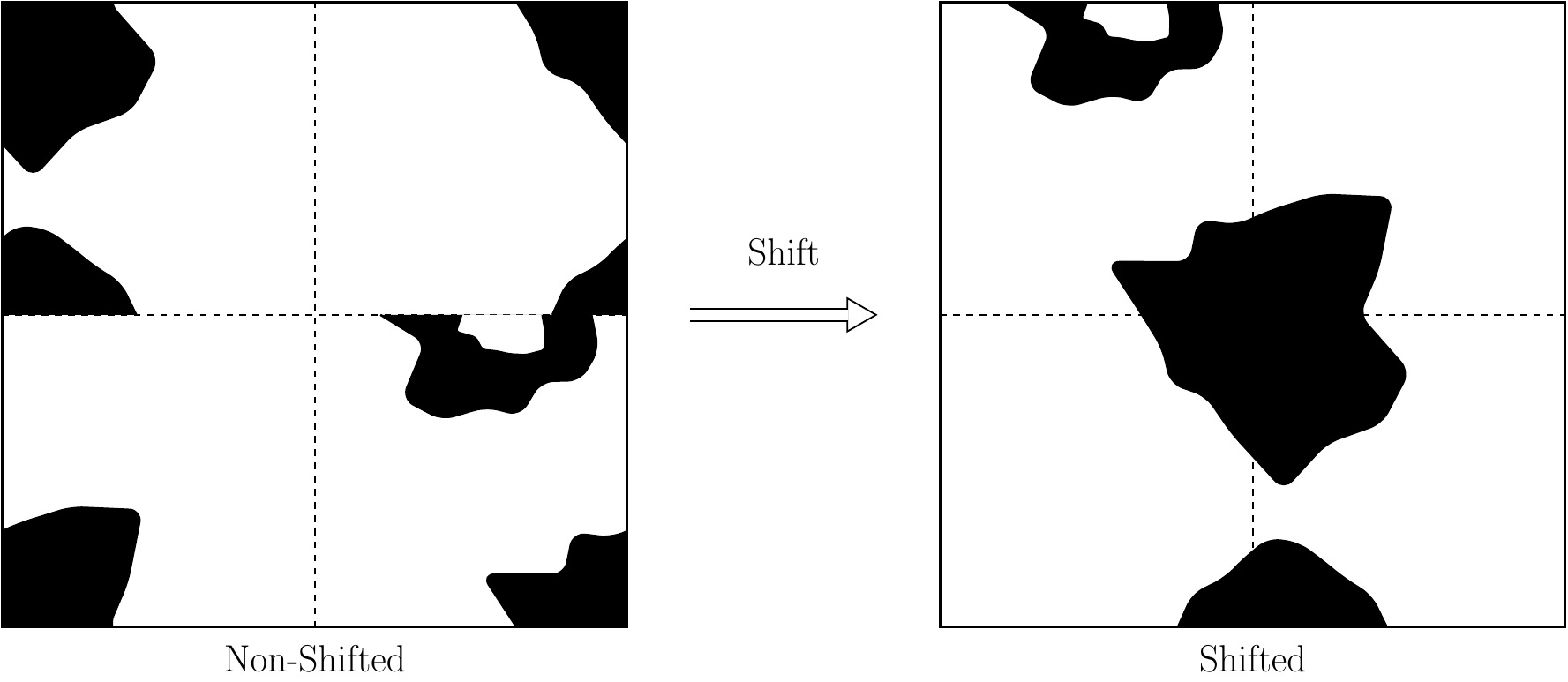}
\caption{Example of thresholded dirty image not shifted after FFT (left image) and its shifted version (right image).}
\label{fig:labelling1}
\end{figure*}

\begin{figure*}
\centering
\includegraphics[width=0.9\textwidth]{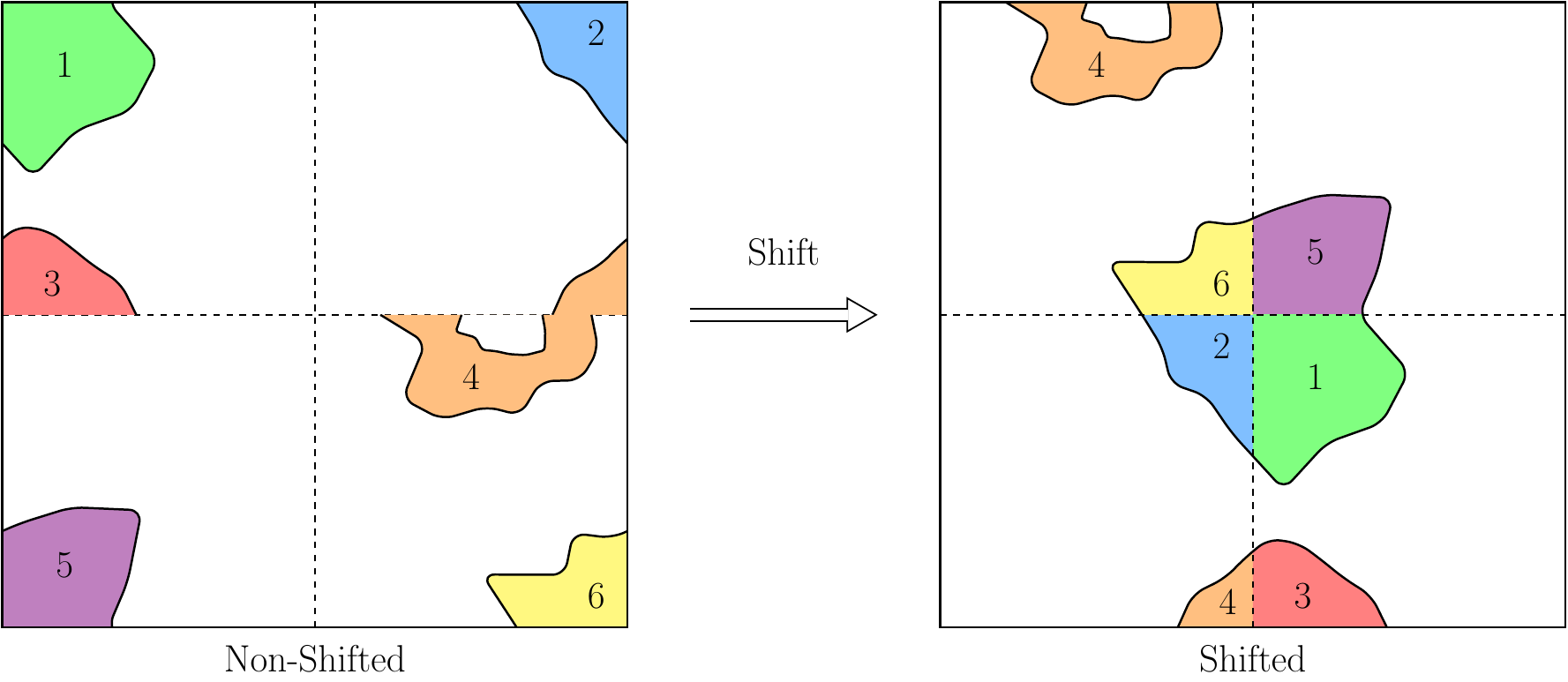}
\caption{Example of label map generated by the original CCL algorithm for non-shifted input (left image) and its shifted version (right image).}
\label{fig:labelling2}
\end{figure*}

\paragraph{Optimisation 1}

Interferometric images are made by Fourier transforming the native measurement data from radio interferometers. Typically this is done using the Fast Fourier Transform (FFT), implementations of which require a quadrant shift to be made on the output data grid in order to correctly align data values. Since this shift requires copying the image data with non-sequential accesses, it tends to be computationally expensive. For example, in the STP, it was observed that FFT quadrant shift may increase the imager running time up to 20\%. In this context, the first optimisation implemented here intends to avoid the need to perform quadrant shifting prior to the source finding process.

An example of a binary image that results from thresholding a non-shifted image is shown in the left image of Figure \ref{fig:labelling1}. The corresponding shifted version is presented in the right image of Figure \ref{fig:labelling1}. If the non-shifted image was used as input to the original CCL function, without the proposed changes, the labelling result would be as illustrated in Figure \ref{fig:labelling2}. Six components would be detected in this example (see the left image of Figure \ref{fig:labelling2}). When the labelled image is shifted the results are not as expected, because multiple labels appear as being assigned to the same object (see right image of Figure \ref{fig:labelling2}).

\begin{figure*}
\centering
\includegraphics[width=0.9\textwidth]{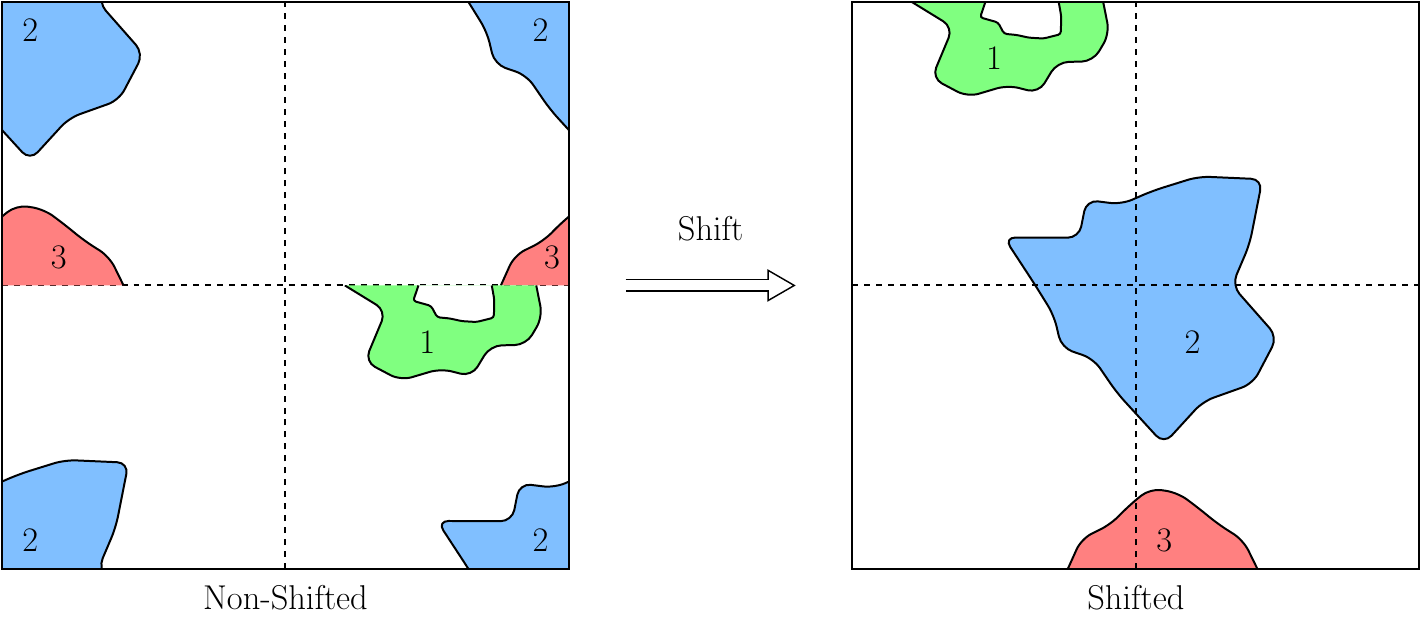}
\caption{Example of label map generated by the proposed CCL algorithm for non-shifted input (left image) and its shifted version (right image).}
\label{fig:labelling3}
\end{figure*}

\begin{figure*}
\centering
\includegraphics[width=0.9\textwidth]{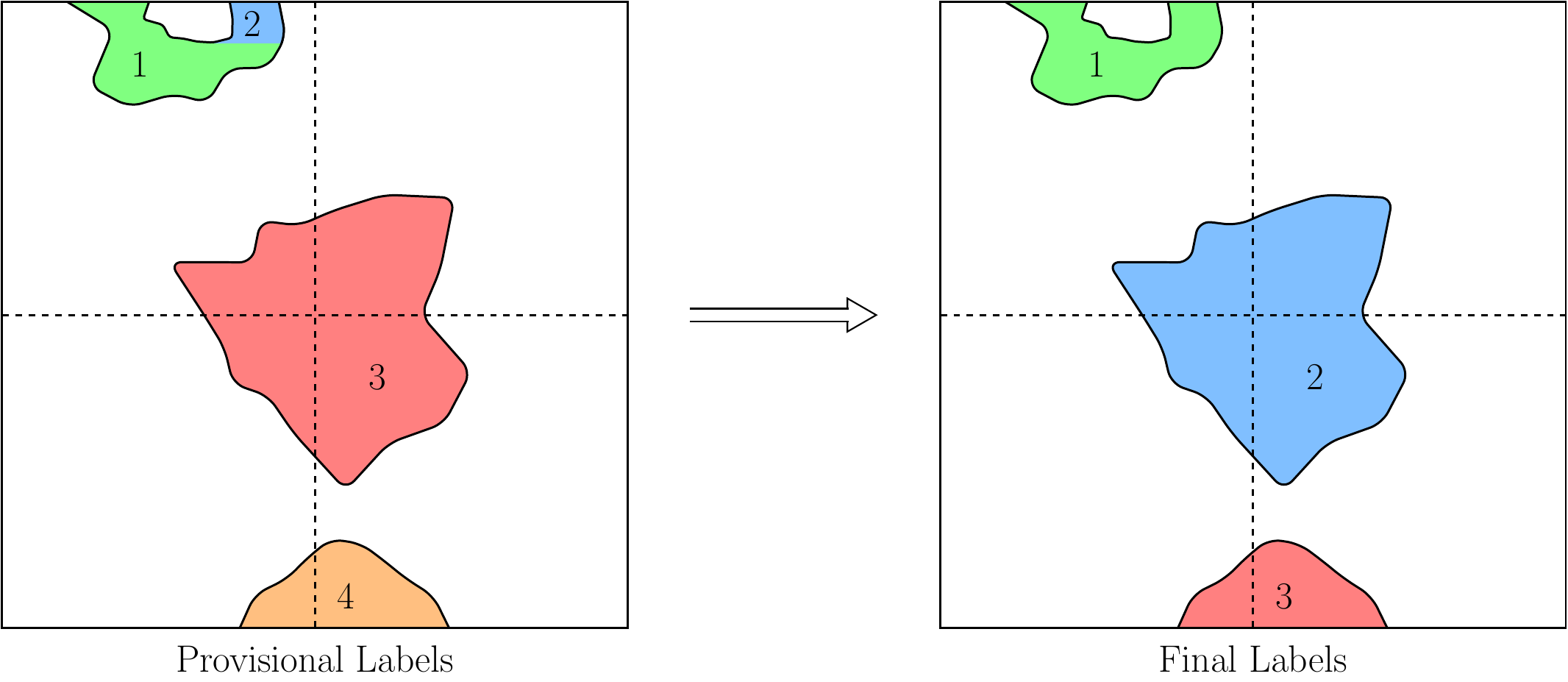}
\caption{Example of label map generated by the original CCL for shifted input image. The left image shows intermediate scanning phase results using the provisional labels and the right image shows the final labelling output using the labels determined in the analysis phase.}
\label{fig:labelling4}
\end{figure*}

To fix this issue, the CCL algorithm has been modified so that it produces a label map as shown in Figure \ref{fig:labelling3}. As can be observed, the shifted version of the label map (right image of Figure \ref{fig:labelling3}) presents only 3 labels as expected.
To explain the proposed change, we first review the phases of the CCL algorithm: the scanning, analysis and labelling. In the scanning phase, provisional labels are assigned to object pixels. In this phase, multiple labels can be assigned to one connected component, resulting in a provisional label map as illustrated in the left side of Figure \ref{fig:labelling4}, that contains 4 provisional labels. After determining the final labels in the analysis phase and performing the final labelling phase, the resulting label map is the expected one, with 3 labels, as illustrated in the right side of Figure \ref{fig:labelling4}.

\begin{figure*}
\centering
\includegraphics[width=0.9\textwidth]{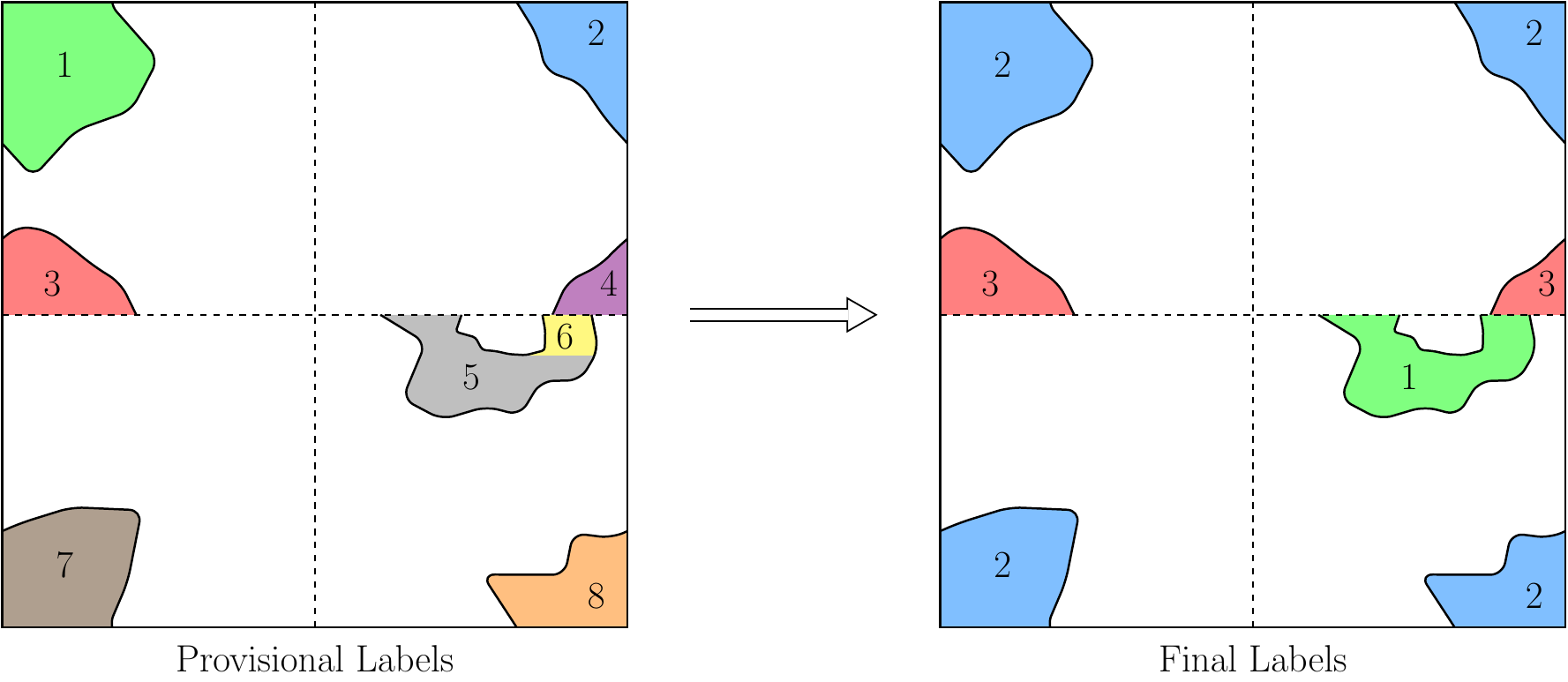}
\caption{Example of label map generated by the proposed CCL for non-shifted input image. The left image shows intermediate scanning phase results using the provisional labels and the right image shows the final labelling output using the labels determined in the analysis phase.}
\label{fig:labelling5}
\end{figure*}

The proposed solution modifies the scanning phase of CCL to assume discontinuities at the horizontal and vertical dashed lines in the middle of the image (see the left image of Figure \ref{fig:labelling5}). Thus, when assigning the provisional labels, different numbers are assigned to samples separated by these lines. An example illustrating the result of the modified scanning phase for the non-shifted image is given in the left side of Figure \ref{fig:labelling5}. In this example, labels 4 and 6 are created due to the horizontal dashed line in the image centre representing a discontinuity.

A second change is related to the label equivalence information created during the scanning phase. An equivalence is defined between provisional labels that are assigned to the same connected component (or object). Since the objects in the non-shifted image can be partitioned at the image margins (see for instance, the object with label 2 in the right image of Figure \ref{fig:labelling5}), an equivalence between provisional labels assigned to objects at the image margins shall be created. In practice, it is added an new step that creates equivalences between provisional labels touching opposite image margins.
This process is denominated as border merging and it is also used on parallel implementations of CCL. After border merging, the analysis and labelling phases are applied, producing the expected result as shown in the right image of Figure \ref{fig:labelling5}.

\begin{figure*}
\centering
\includegraphics[width=0.9\textwidth]{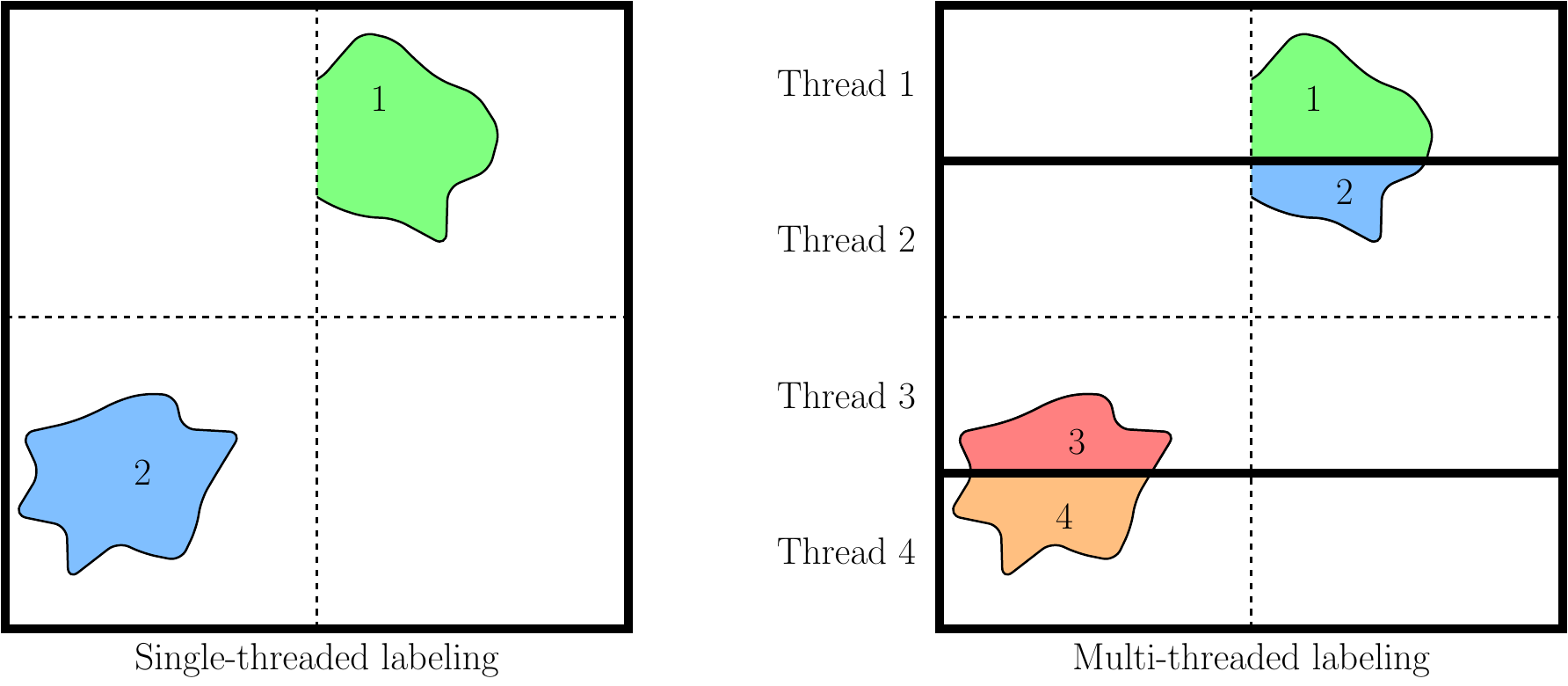}
\caption{Single-threaded (left) vs multi-threaded (right) implementation of the proposed CCL function. Border merging is applied over the thick solid lines.}
\label{fig:parallel_labelling}
\end{figure*}

It is important to refer that the changes introduced by this optimisation may slightly increase the computational complexity of the labelling stage, typically up to a maximum of 3\%. This increase is mostly due to the additional conditions required in the scanning phase. The image data continues being scanned sequentially in the memory thus avoiding a high performance penalty.

\paragraph{Optimisation 2}

The second proposed improvement is the use of parallel processing to reduce the running time of CCL. The proposed algorithm uses a multi-threaded implementation of CCL, that parallelises the scanning and labeling phases. Parallelisation of the analysis phase is not possible due to the way the equivalence information is processed to determine the final labels. This is not a significant issue as the analysis phase presents a low complexity, especially when the number of provisional labels is not significant.

For an efficient parallel processing, the image should be partitioned in horizontal or vertical strips, depending whether the image is stored in row-major or column-major order, and each thread assigned to a distinct strip. Other partitioning approaches, like the use of quadrants are not recommended and would create a high performance penalty, since the threads would be assigned to non-contiguous blocks of data in the system's memory. 

While the labeling phase can be easily implemented in parallel by assigning different strips of the binary image to each thread, the scanning phase requires some extra work.
For the scanning phase, the image also can be divided into strips that are assigned to different threads. However, this procedure changes the labeling results of the algorithm, specifically when the one connected component is defined across two strips. In that case, different provisional labels are assigned by each thread to each partition of the object. An example is illustrated in the right side of Figure \ref{fig:parallel_labelling} for a image represented in row-major order. This issue can be resolved by performing border merging over the horizontal strip borders, with the exception of the dashed lines in the middle of the image.
Figure \ref{fig:parallel_labelling} shows the border merging lines in the single-threaded and multi-threaded implementations represented by the thick solid lines. 

\paragraph{Optimisation 3}

The third proposed optimisation is to perform the image thresholding during the scanning phase of CCL. Instead of generating a new array to store the binary image, the CCL function receives the original dirty image as input. When it scans the image, each sample is compared with the analysis threshold, the binary value is determined and the provisional label is generated.

\paragraph{Optimisation 4}

The fourth proposed optimisation merges the last step of valid label detection with the labeling phase of CCL. For detecting valid labels, the maximum sample value of each object or source is compared with the detection threshold. 
A positive source that contains at least one sample above the positive detection threshold is considered valid. A negative source is valid only if it contains at least one sample below the negative detection threshold.
Merging this procedure with the last phase of the CCL allows one to avoid an extra pass over the label map and thus to reduce the number of memory accesses. Its implementation uses parallel processing, performed in the same loop as the labeling phase of CCL.

\paragraph{Optimisation 5} 

As a final optimisation, the proposed CCL function was designed to perform the search for positive and negative sources at the same time. This optimisation will avoid calling the CCL function two times for each of both positive and negative sources, which would result in four passes over image data. The proposed solution only requires two passes to label both positive and negative sources. To make it possible, we use two structures to independently store the equivalence information for positive and negative sources. An advantage of this solution is that the same label map can be used to represent the positive and negative sources, as they do not overlap. This allows one to significantly reduce the amount of memory usage as well as the number of memory reads and writes. We note that the alternative is to perform two runs of CCL to detect the positive and negative sources and then combine the two resulting label maps.

\subsubsection{Data Passes}

The optimisations described above result in a significant improvement of the performance of the detection algorithm for positive and negative sources, reducing the number of image passes to two, which is the minimum required for the CCL algorithm. 
After detection of the valid sources, the algorithm computes the first and second statistical moments of each valid source. To compute these moments, a third pass over the label map is required.
In this pass, some auxiliary data required for the Gaussian fitting step is also computed, namely the bounding box and the number of samples for each region.
Optionally, the third pass may be also used to remove invalid sources (those below the \emph{detection} threshold) from the label map. The final label map may be desired for instance for debugging purposes, \emph{e.g.} to be observed and analysed by the user.

\begin{landscape}

\begin{table*}
\centering
\caption{Main functions of the labeling procedure and the required number of image passes for two implementation approaches: straightforward and optimised. 
An image pass implies writing operations on the image unless otherwise stated, \emph{e.g} ``only read''.}
\label{tab:labelling_passes}
\begin{tabular}{@{}lcc@{}}
\toprule
\multicolumn{1}{c}{}                & \multicolumn{2}{c}{\textbf{No of image passes}} \\ \cline{2-3} \noalign{\vskip 0.1cm}
\textbf{Labelling functions}                   & \textbf{Straightforward approach} & \textbf{Optimised approach}   \\ \midrule
Positive thresholding               & 1                   & \multirow{4}{*}{1}    \\
Scanning phase - positive sources   & 1                   &                       \\
Negative thresholding               & 1                   &                       \\
Scanning phase - negative sources   & 1                   &                       \\ \hline
Labelling phase - positive sources   & 1                   & \multirow{4}{*}{1}    \\
Detection of valid positive sources & 1 (only read)       &                       \\
Labelling phase - negative sources   & 1                   &                       \\
Detection of valid negative sources & 1 (only read)       &                       \\ \hline
Removal of invalid positive sources & 1                   & \multirow{4}{*}{1}    \\
Removal of invalid negative sources & 1                   &                       \\
Combination of label maps           & 1                   &                       \\
Computation of source barycentre    & 1 (only read)       &                       \\ \hline
\textbf{Total number of image passes} & \textbf{9 + 3 (only read)}       & \textbf{3}            \\ \bottomrule
\end{tabular}
\end{table*}

\end{landscape}

Table \ref{tab:labelling_passes} summarises the major steps of the labeling algorithm and the number of required passes over the image (which may refer to the dirty image or label map), specifically for two implementation approaches: the straightforward and the optimised. 
An image pass implies write operations on the image unless otherwise stated, \emph{e.g} ``only read''.
The straightforward approach refers to a trivial implementation of the labeling function, while the optimised approach corresponds to the proposed solution that minimises the number of data writings and readings.

It can be observed that the proposed optimised approach requires only 3 image passes while the straightforward implementation uses 12 image passes, although not all of them involve write operations. Such a large reduction in the number of passes, relative to the straightforward implementation, results in a significant performance improvement for the source finding.
We do not believe that it is possible to reduce the number of passes further due to the data dependencies in the processing.

In the first image pass of the optimised approach the thresholding step and the scanning phase of CCL are performed for both positive and negative sources. 
The second image pass of the optimised solution is used for the final labeling phase of CCL and the detection of valid sources. 
Finally, the third pass is used to compute the statistical moments of each source, compute auxiliary data about each source and optionally to update the label map, removing any invalid sources. 
The optimised solution does not need to combine the label maps of negative and positive sources, since the labeling function directly generates a combined source map.

\subsection{Gaussian fitting}
\label{sec:gauss}

As a final step in the source finding algorithm, Gaussian functions are fitted to each detected source or island. 
This procedure consists of solving non-linear least squares problems to estimate the parameters of the Gaussian functions that fit  each island. Gaussian fitting has been implemented using the Ceres Solver library \cite{ceressolver}.

The Ceres Solver first models the problem and then solves it. For modeling, the Ceres Solver provides three main methods to compute the derivatives:
\begin{itemize}
 \item Analytic: The derivatives shall be manually implemented by the user.
 \item Numeric: Ceres Solver numerically computes the derivatives using finite differences. 
 \item Automatic: Ceres Solver automatically computes the analytic derivatives using C++ templates and operator overloading. 
\end{itemize}

In the algorithm developed, analytic or automatic derivatives can be used, depending on the chosen option in the input configuration file to the program. We did not implement numeric derivatives because they present a larger computational complexity. In this work it was observed that the analytic derivatives tended to perform more efficiently than automatic derivatives.

For the solver step, the Ceres Solver provides several methods. Through the configuration file, it is possible to choose between the trust region and line search minimiser methods to solve the non-linear least squares problem. 
The trust region method approximates the objective function using a model function (often a quadratic) over a subset of the search space known as the trust region. By default, we use the Levenberg-Marquardt algorithm and the dense QR factorization routines of the Eigen library to solve the trust region problem. 
For the line search approach, the BFGS or LBFGS method may be used to choose a search direction. These are a generalisation of the conjugate gradient method for non-linear functions.

The Gaussian fitting step computes the following list of parameters for each detected island:
\begin{description}
 \item[amplitude] (double) the amplitude parameter of the Gaussian function;
 \item[x\_center, y\_center] (double) the x, y coordinates of the Gaussian function centre in the image;
 \item[semimajor] (double) the length of the major Gaussian axis in pixels;
 \item[semiminor] (double) the length of the minor Gaussian axis in pixels;
 \item[theta] (double) the position of the semimajor axis measured counter-clockwise from the x axis (in radians).
\end{description}

In addition to the Gaussian parameters derived by least-squares optimisation, an initial Gaussian estimation based on the first and second statistical moments is outputted. Thus, each detected island is accompanied by two sets of Gaussian parameters. Jointly, the following data about each detected island is provided:
\begin{description}
\item[sign]  (integer) represents whether the transient source is positive or negative. Possible values are $+1$ and $-1$;
\item[val] (double) represents the extremum (maximum or minimum) pixel value for the (positive or negative) transient source;
\item[x\_idx, y\_idx] (integer) represent the x, y index values (\textit{i.e.} pixel coordinate in the image) of the extremum value;
\item[num\_samples] (integer) indicates the number of pixels in the island;
\end{description}

In terms of computational complexity, the Gaussian fitting step usually does not present a significant impact on the overall running time of the pipeline, especially when the number of detected islands is not significant. However, when the number of detected islands increases the computational cost of this step may be noticeable.


\section{Tests and benchmarks}
\label{sec:tests_benchmarks}

Here we present and analyse the benchmark results of the main functions that constitute the source finding algorithm developed.
Benchmarks were performed on an High Performance Computer (HPC) with two processor sockets, specifically the Intel Xeon CPU E5-2650 v4 working at 2.20GHz and using 504 GB of RAM. Each CPU has 12 physical cores that provide 24 threads due to Hyper-Threading (HT) technology, summing a total of 48 threads.

The HT technology doubles the number of available threads in a multi-core CPU by duplicating certain parts of the processor which allows each processor core to appear as the usual physical core and an extra logical core to the operating system.
While hyper-threading tends to provide a superior computing performance for certain cases, the parallel efficiency of each thread tends to be inferior due to hardware resources of one physical core being shared between two threads. For this reason, this technology has been disabled for the performed experiments. The amount of threads used in the HPC was thus 24.

The benchmarking tests discussed here are based on the measured running times, which provide a rough idea of the relative computational complexity of the tested functions. It is important to note that these timings may vary between distinct benchmarking instances, even on the same hardware, as they may be affected by the operating system tasks running concurrently and as well as its status.

The following experiments are presented and discussed in this section:
\begin{itemize}
 \item Median - evaluate the running time of the proposed median functions discussed in \S~\ref{sec:backest} for different sized matrices of random data, comparing with other median implementations, as well as the single- and multi-threaded execution. 
 \item Standard deviation - evaluate the running time of the standard deviation function proposed in \S~\ref{sec:rms} for different sized matrices of random data, comparing with other standard deviation implementations, as well as the single- and multi-threaded execution. 
 \item Steps of source find - evaluate the running times of the main steps of the source find algorithm proposed here using simulated astronomical images.
 \item Source find - evaluate the total running time of the source find algorithm proposed here using simulated astronomical images and comparing single- and multi-threaded executions.
\end{itemize}

The simulated astronomical data used for source find testing contains 3 gaussian-shaped sources. Such reduced number of sources was used because we are using residual images generated in the context of the STP development. There is not any limitation in the algorithm relative to the number of sources present in the image. In fact, we used a larger number of sources in the experiments of \S~\ref{sec:comparisons} where the algorithm is compared with other source finders.

\begin{figure}[t]
\centering
\includegraphics[width=\textwidth]{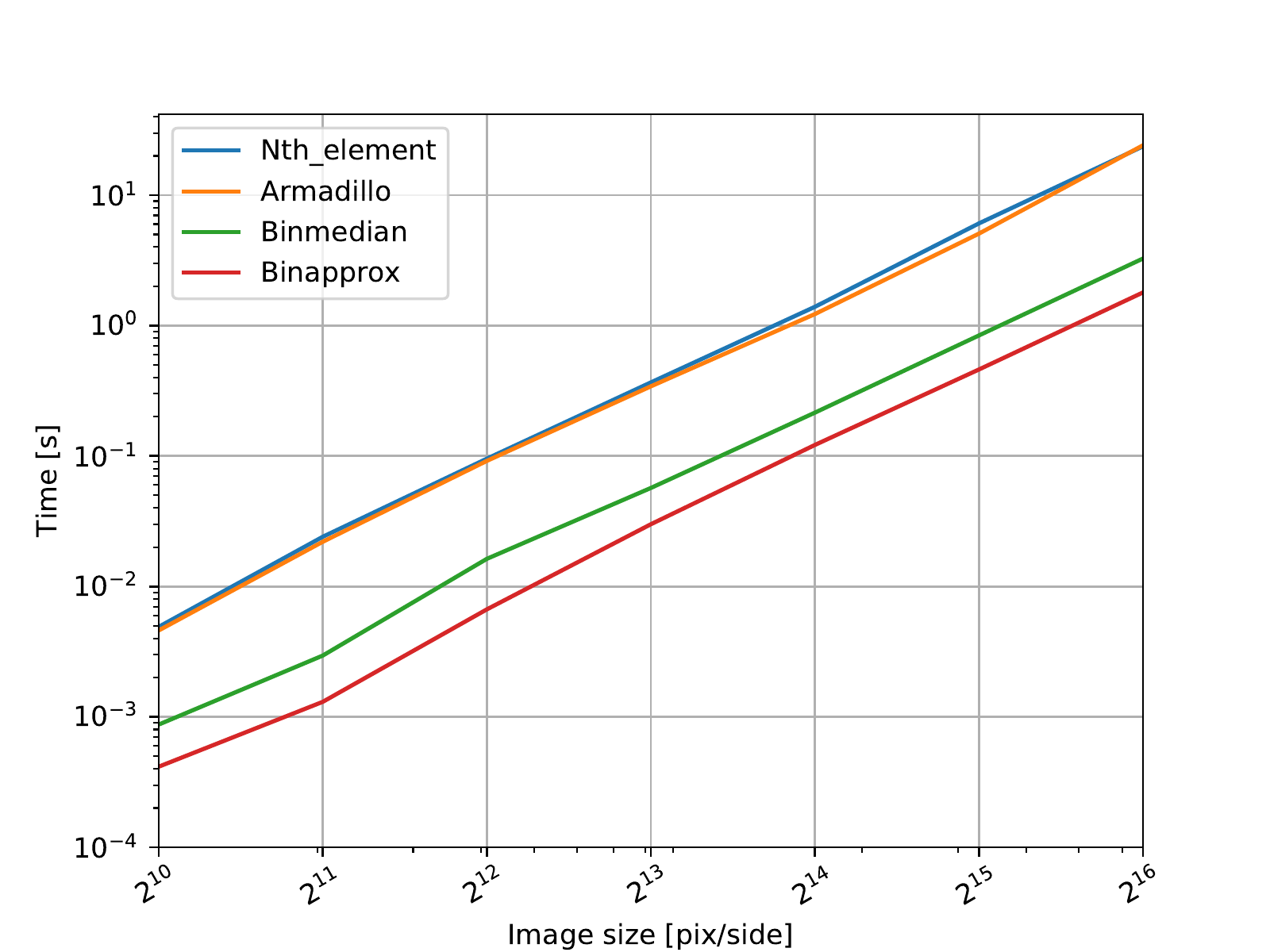}
\caption{Running times (in seconds) for computing the median using different image sizes and the following methods: a method based on the \emph{Nth\_element} function, the Armadillo library, the \emph{binmedian} and the \emph{binapprox} methods.}
\label{fig:times_median_multithread}
\end{figure}

\begin{figure}[t]
\centering
\includegraphics[width=\textwidth]{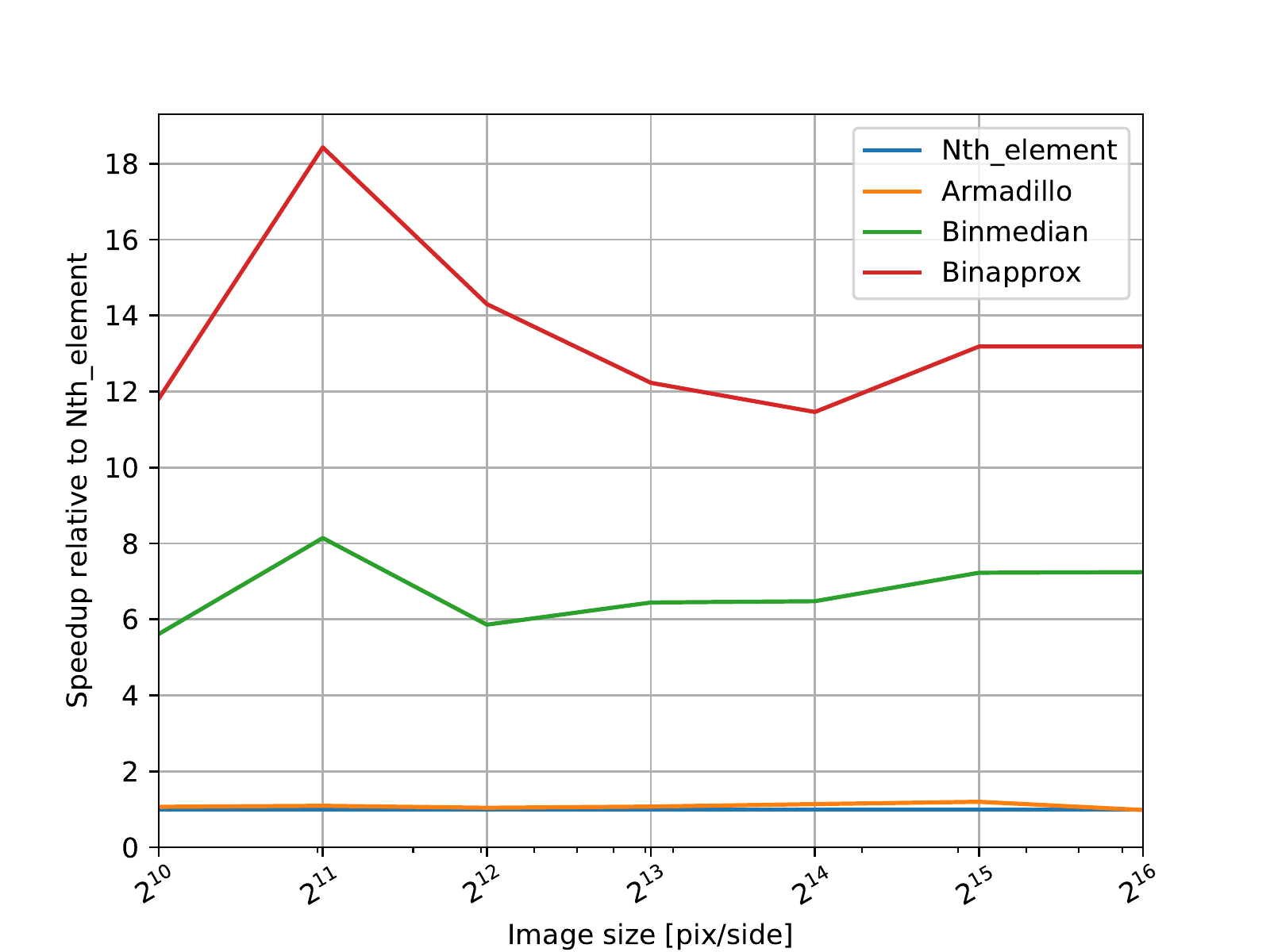}
\caption{Speed up gains of the tested median functions for different image sizes, namely one based on the Armadillo library, the \emph{binmedian} and the \emph{binapprox} methods, relative to the reference method based on the \emph{Nth\_element} function.}
\label{fig:speedup_median}
\end{figure}

\begin{figure}[t]
\centering
\includegraphics[width=\textwidth]{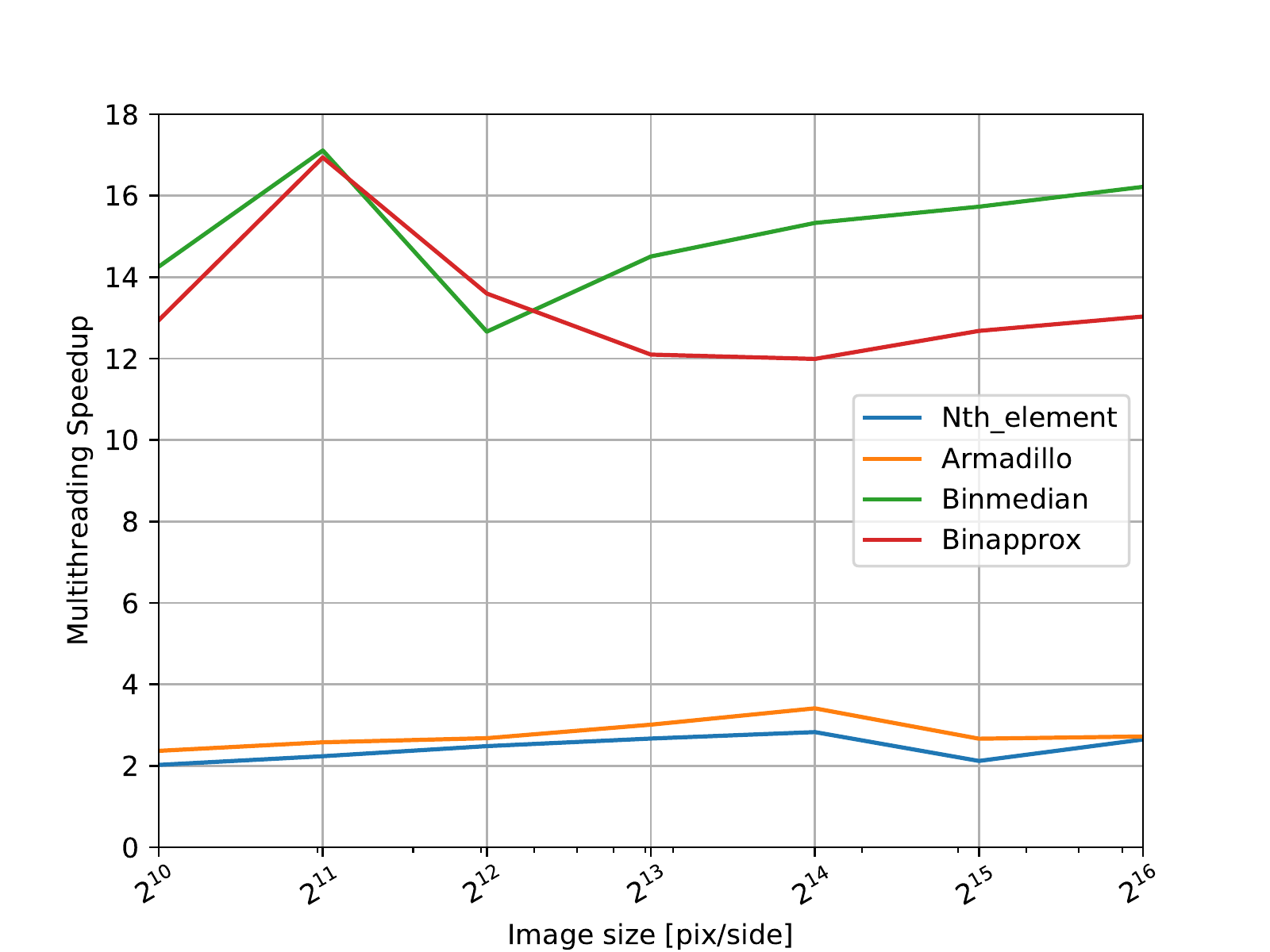}
\caption{Multithreading speed up gains of the tested median functions for different image sizes: a method based on the \emph{Nth\_element} function, the Armadillo library, the \emph{binmedian} and the \emph{binapprox} methods.}
\label{fig:speedup_median_multithread}
\end{figure}

Figure \ref{fig:times_median_multithread} presents the running time of different implementations of the median function used for background level estimation, specifically:
\begin{itemize}
 \item \emph{Nth\_element} method - provided in the C++ standard library for $k^{th}$ smallest element selection based on \emph{quickselect} algorithm;
 \item \textit{arma::median} method of the Armadillo library \cite{Curtin_JOS2016} - a high quality linear algebra library for the C++ language used in this work for matrix representation. It aims towards a good balance between speed and ease of use;
 \item \emph{binmedian} method - proposed in \S~\ref{sec:backest};
 \item \emph{binapprox} method - proposed in \S~\ref{sec:backest};
\end{itemize}

The respective speed up gains, using the \emph{Nth\_element} function as reference method are given in Figure \ref{fig:speedup_median}. 
From these results, it may be concluded that the \emph{binapprox} method provides the best performance, presenting speed up gains between 12 and 18 relative to the reference \emph{Nth\_element} function. 
The exceptional running performance for the $2^{11}\times 2^{11}$ image size can be related with the cache memory optimisation. As such small image sizes mostly fit into the cache memory, the associated execution performance is easily affected by the cache optimisation, which depends on the hierarchical cache sizes, number of threads and allocated buffer sizes.

Despite its superior performance, it is important to note that the \emph{binapprox} method does not compute the exact median value. If exact median is required, the \emph{binmedian} method provides the best results, presenting a speed up gain between 6 and 8 relative to the \emph{Nth\_element} function. 
Regarding the method provided by Armadillo, its performance is close to the one that uses the \emph{Nth\_element} function. This is not a surprising result since Armadillo's implementation of median is based also on the \emph{Nth\_element} function.

It is important to remember that although all the tested algorithms shown in Figures \ref{fig:times_median_multithread} and \ref{fig:speedup_median} use multi-core processing, they do not present the same parallel efficiency. To estimate the speedup gains provided by parallel implementation using TBB, the previously presented median functions were run in single-threaded mode and the speedup gains were computed.
The achieved multi-threading speedup gains are shown in Figure \ref{fig:speedup_median_multithread}. 
It can be observed that the methods based on the \emph{Nth\_element} function (which includes Armadillo) present a multi-threading speedup between 2 and 3, corresponding to a small parallel efficiency of between 8\% and 12.5\% in the 24-core HPC. 
These methods use a parallel version of the \emph{Nth\_element} function provided by the Standard C++ Library, which does not scale efficiently due to the characteristics of its algorithm based on \emph{quickselect}. 
Differently, the \emph{binmedian} and \emph{binapprox} methods benefit significantly from parallel processing, presenting speedup gains between 12 and 17, which correspond to a parallel efficiency between 50\% and 70\% in the used computer. 

In fact, the good performance of the \emph{binmedian} and \emph{binapprox} methods is possible due to the high parallel efficiency of these functions. When using a single-core system, the advantage of the \emph{binmedian} method over the Armadillo median function is irrelevant. In some cases, the single-threaded version of \emph{binmedian} can be even worse than the median methods based on the \emph{Nth\_element} function. The use of multi-core processing is thus the key to take advantage of the \emph{binmedian} and \emph{binapprox} algorithms. 

\begin{figure}[t]
\centering
\includegraphics[width=\textwidth]{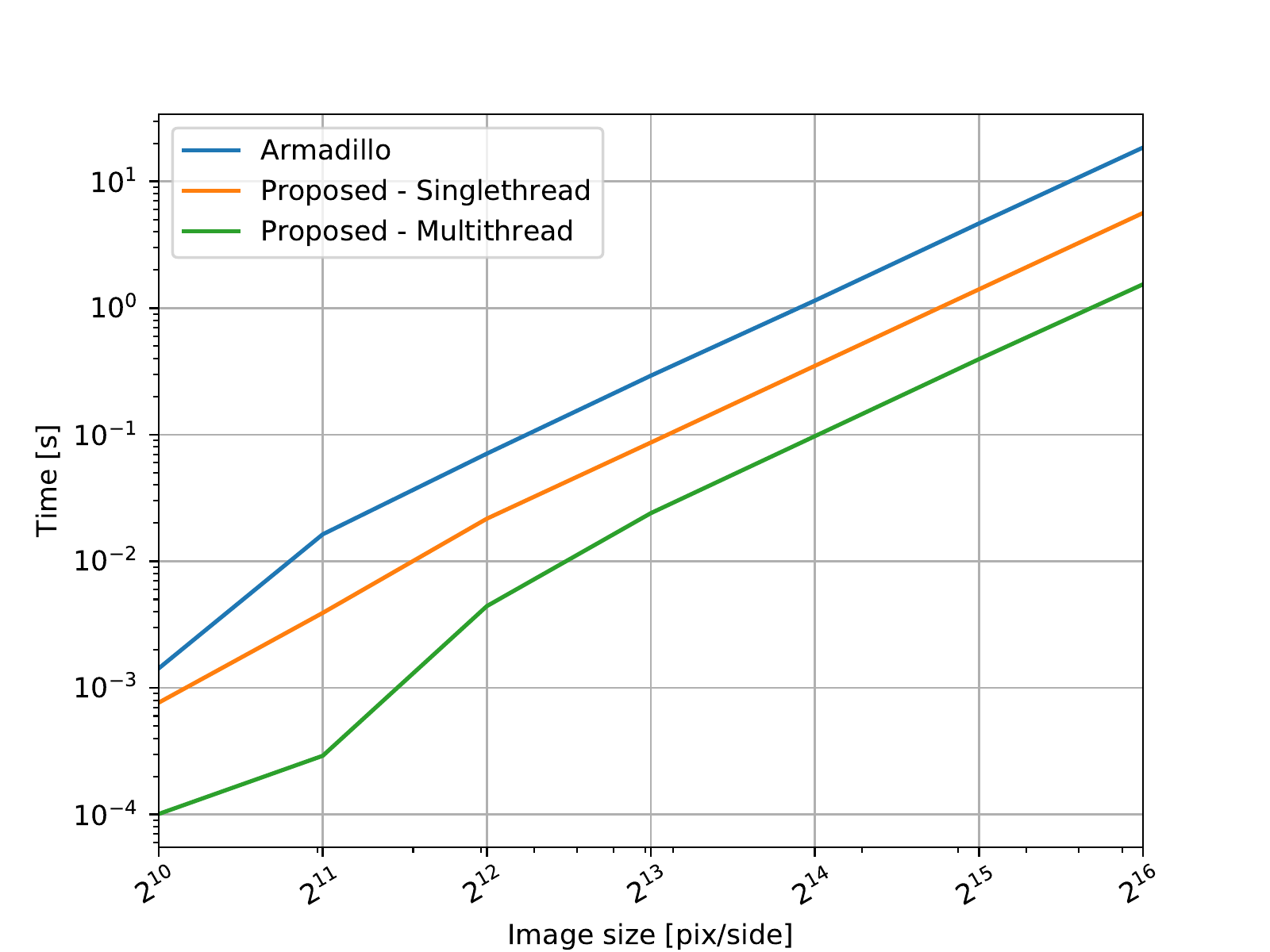}
\caption{Running times (in seconds) of the standard deviation function using the method provided by Armadillo and the proposed TBB-based multi-threaded implementation, for different image sizes.}
\label{fig:times_stddev}
\end{figure}

\begin{figure}[t]
\centering
\includegraphics[width=\textwidth]{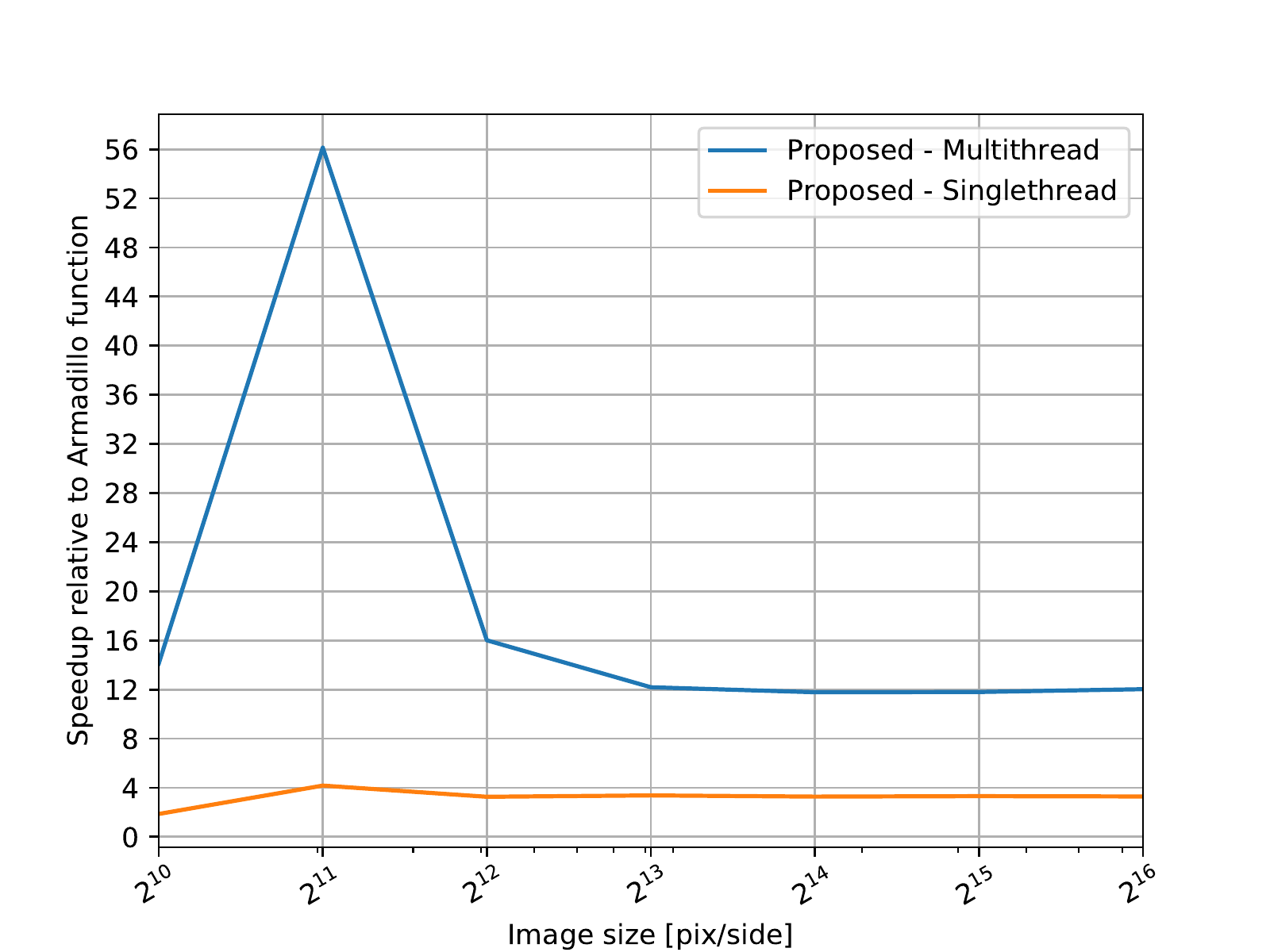}
\caption{Speed up gains of the proposed standard deviation function using the single-threaded and TBB-based multi-threaded implementations, computed relative to the reference \textit{stddev} method provided by Armadillo, for different image sizes.}
\label{fig:speedup_stddev}
\end{figure}

Another important function of STP, that is used also as a step of the \emph{binmedian} and \emph{binapprox} methods is the standard deviation function.
Figure \ref{fig:times_stddev} presents the running times of two implementations of standard deviation function, namely the one provided by Armadillo library and the proposed solution based on TBB (see \S~\ref{sec:rms}) executed in both single-threaded and multi-threaded modes. The speed up results of the proposed function relative to the Armadillo method are presented in Figure \ref{fig:speedup_stddev}. It can be observed that the proposed TBB-based solution presents significant speed up gains, being approximately 4 and 12 times faster than Armadillo for the single- and multi-threaded modes, respectively. For the smaller image sizes, the speed up results present a larger variation, due to the fact that these sizes mostly fit into the cache memory, being easily affected by the cache optimisation techniques.

Usually the mean quantity is computed as the first step of the standard deviation function. However, the proposed function uses an alternative approach that avoids the mean computation. The developed function computes two accumulation values as explained in \S~\ref{sec:rms} using a single pass over the array. These accumulation values are then used to derive both the mean and standard deviation of the data. The fact that the developed function provides both these quantities at once is another important advantage, since it avoids the necessity to call the mean function when both the mean and standard deviation quantities are required. 

\begin{figure*}
\centering
\begin{subfigure}[b]{0.52\textwidth}
\includegraphics[width=\textwidth]{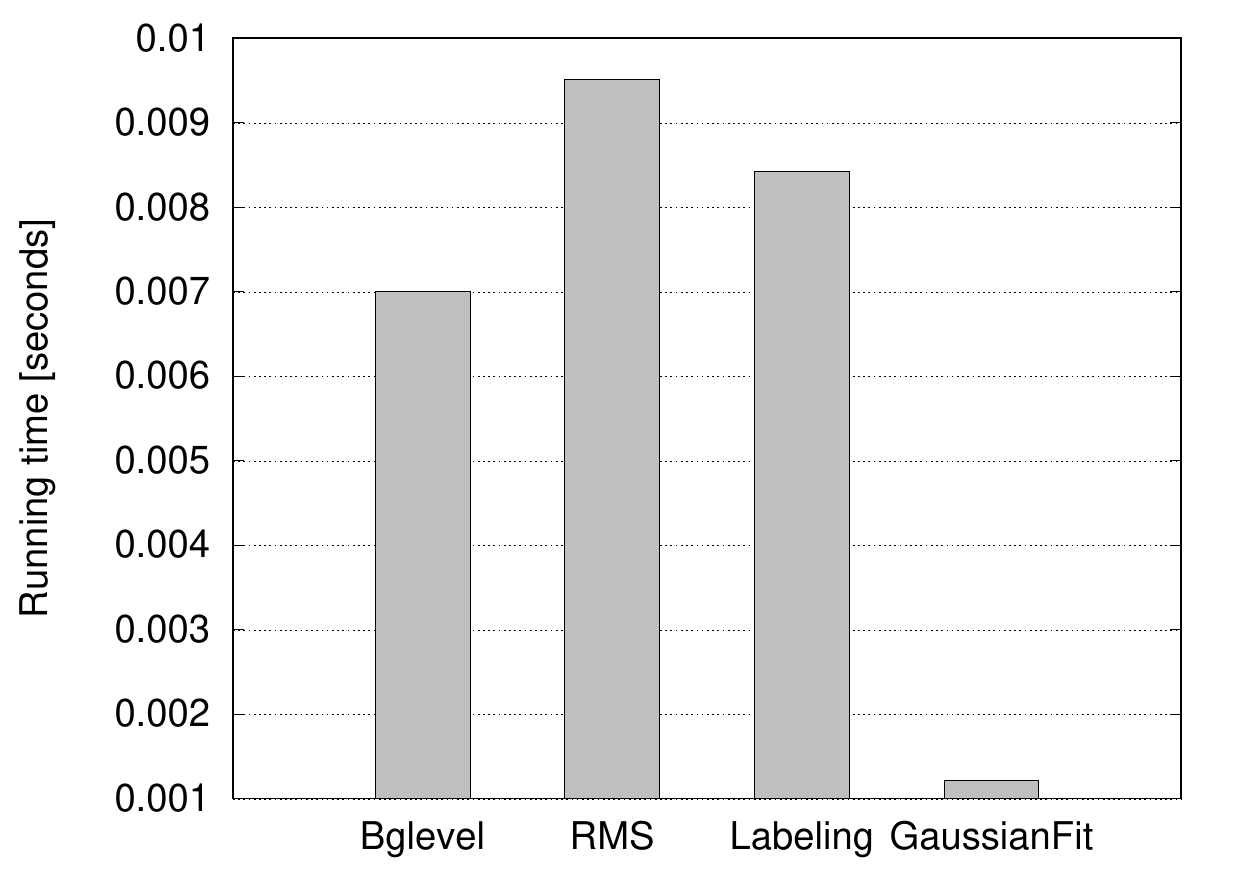}
\caption{Image size = $2^{12} \times 2^{12}$}
\end{subfigure}
\begin{subfigure}[b]{0.52\textwidth}
\includegraphics[width=\textwidth]{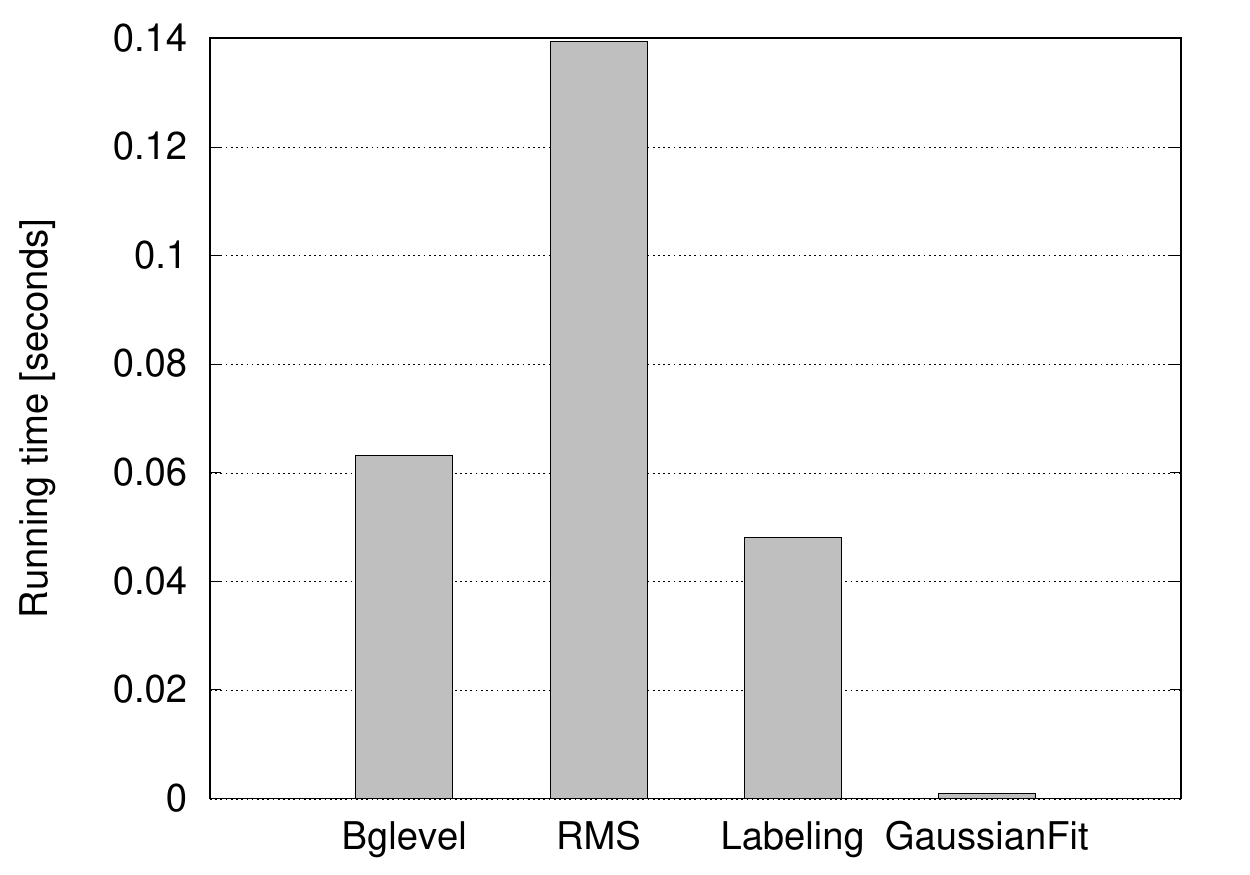}
\caption{Image size = $2^{14} \times 2^{14}$}
\end{subfigure}
\begin{subfigure}[b]{0.52\textwidth}
\includegraphics[width=\textwidth]{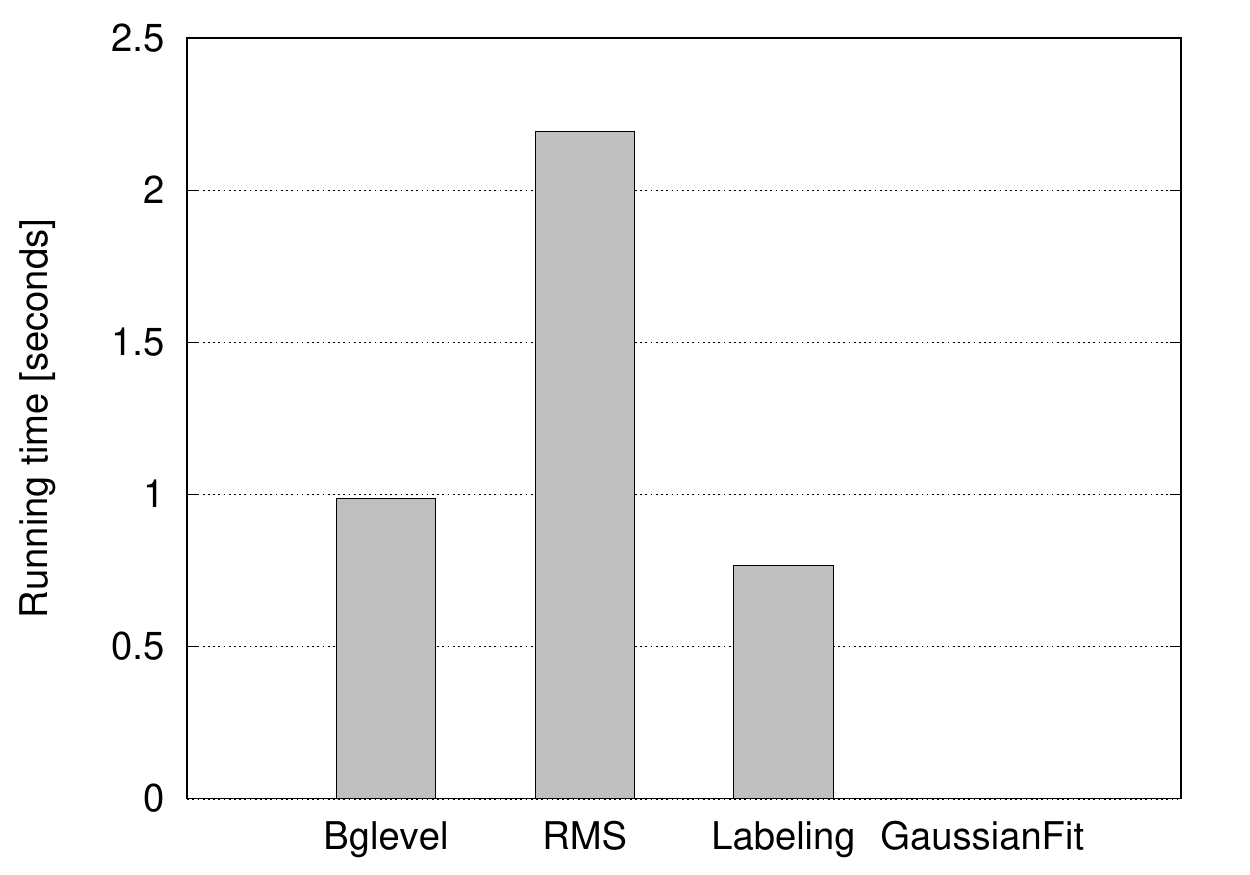}
\caption{Image size = $2^{16} \times 2^{16}$}
\end{subfigure}
\caption{Benchmarking results of the source finding algorithm using different image sizes.}
\label{fig:profiling}
\end{figure*}

Figure \ref{fig:profiling} presents the benchmarking results of the full source finding algorithm, performing the background level estimation (based on \emph{binapprox} function), RMS estimation (with sigma clipping using 5 iterations), source labelling and gaussian fitting. We used simulated images with sizes of $2^{12} \times 2^{12}$, $2^{14} \times 2^{14}$ and $2^{16} \times 2^{16}$.
As can be observed in the referred figures, the gaussian fitting step tends to use an insignificant running time. This is mostly because the test-data has few sources. 
Its timing shall increase in the presence of a larger number of sources.

The most complex step of the source finder is the RMS estimation that needs to perform 5 passes over the image to perform sigma clipping. Due to the optimisations described in this paper, the labelling step performs quite efficiently, presenting execution times inferior to both the background level and RMS estimation functions, for most image sizes. 
When comparing the results for different image sizes, one observes that the relative computational cost of each step does not vary significantly, although the respective absolute timings increase for larger images sizes as expected.

\begin{figure}[t]
\centering
\includegraphics[width=\textwidth]{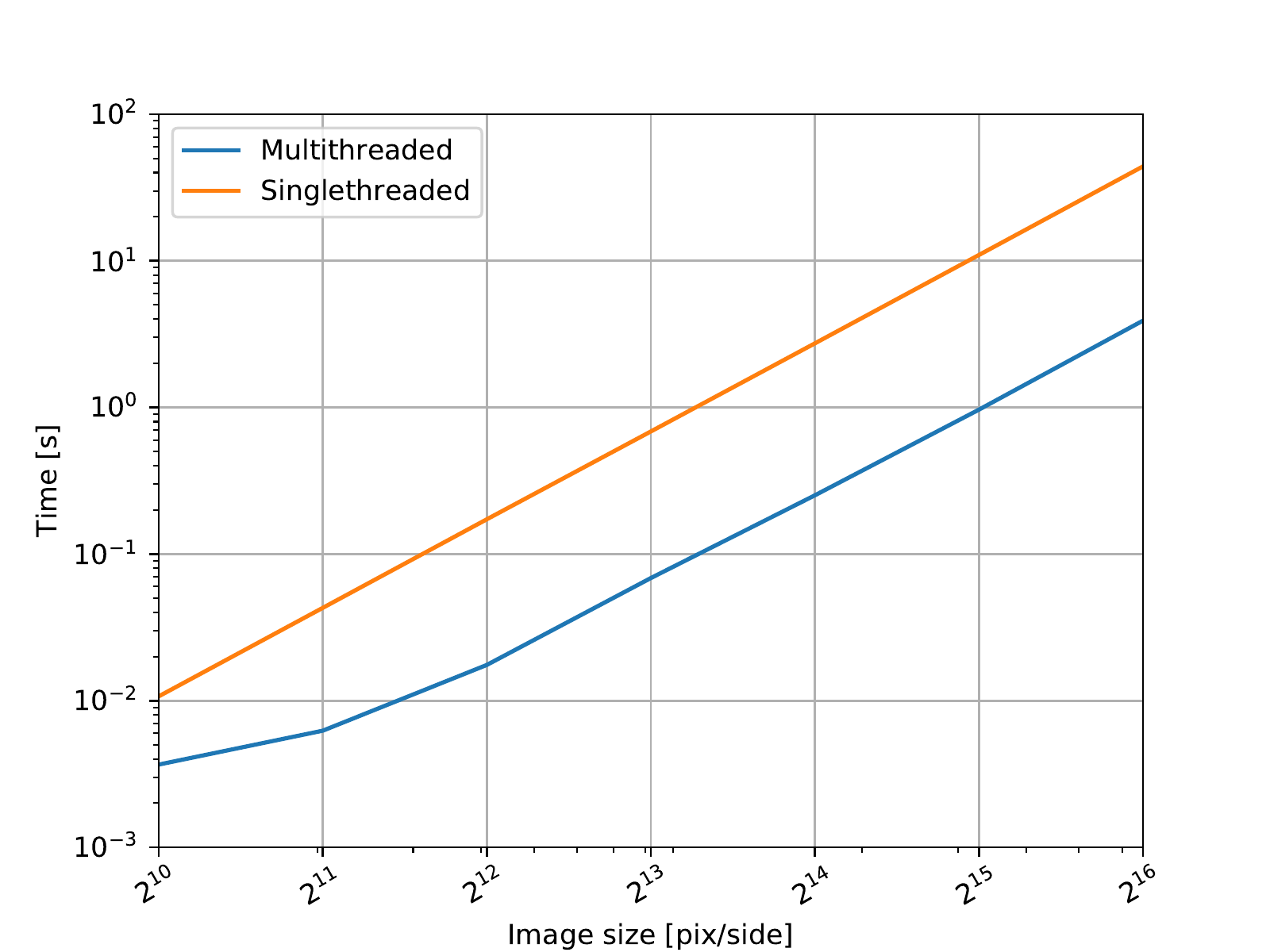}
\caption{Running times (in seconds) of the single-threaded and multi-threaded implementations of the source finding algorithm, for different image sizes.}
\label{fig:times_sourcefind}
\end{figure}

\begin{figure}[t]
\centering
\includegraphics[width=\textwidth]{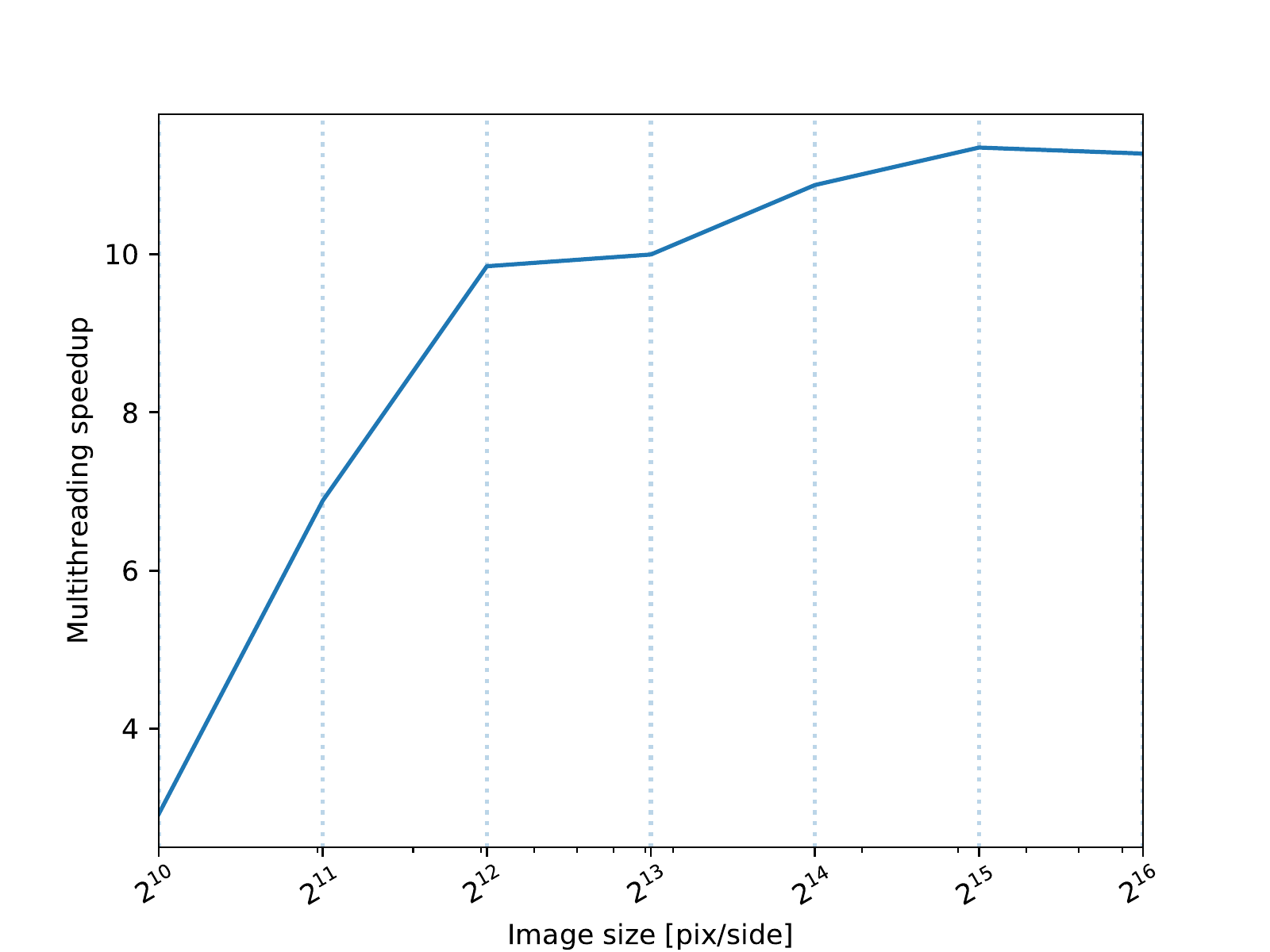}
\caption{Speed up gains of the source finder multi-threaded implementation relative to the single-threaded version, for different image sizes.}
\label{fig:speedup_sourcefind}
\end{figure}

To analyse the parallel performance of the source finding algorithm, Figure~\ref{fig:times_sourcefind} presents the running times of both the single-threaded and multi-threaded execution of the algorithm. The running times of the multi-threaded algorithm vary between 4 milliseconds and 4 seconds, while the single-threaded algorithm varies between 10 milliseconds and 44 seconds, for the image sizes $2^{10} \times 2^{10}$ and $2^{16}\times 2^{16}$, respectively. As can be observed in Figure~\ref{fig:times_sourcefind}, these running times tend to increase exponentially with the image width, increasing linearly with the number of image samples.

The performance gains of the multi-threaded implementation relative to the single-threaded one vary between 3 and 11, as shown in Figure~\ref{fig:speedup_sourcefind}, corresponding to a parallel efficiency of 12.5\% and 46\%, respectively. The multi-threading gains are mostly achieved by the parallel implementation of the median (\emph{binapprox} method), RMS estimation and labelling functions, being more noticeable for larger image sizes.
The fact that the smaller images better fit into CPU cache memory may reduce the parallel efficiency of the algorithm, for instance, due to the false sharing issue caused when multiple threads write data into the same cache line.

\section{Comparative benchmarks}
\label{sec:comparisons}

\begin{figure}[!htb]
\centering
\includegraphics[width=\textwidth]{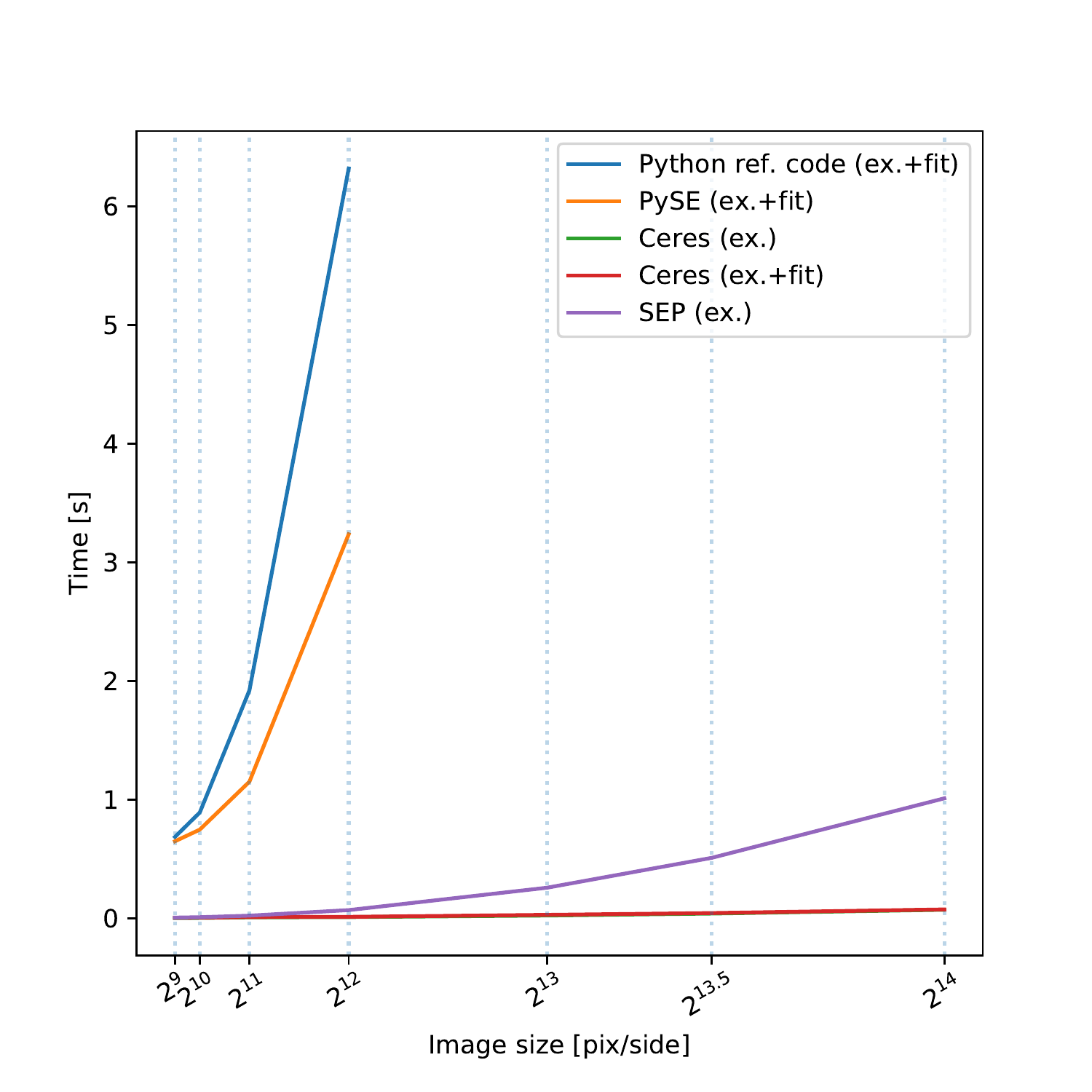}
\caption{Comparative benchmarks for extraction of 64 sources on images of varying size - all sourcefinders.
Some experiments perform only source-extraction (ex.), while others perform both source-extraction and fitting (ex.+fit).}
\label{fig:comparative_extraction}
\end{figure}

\begin{figure}[!htb]
\centering
\includegraphics[width=\textwidth]{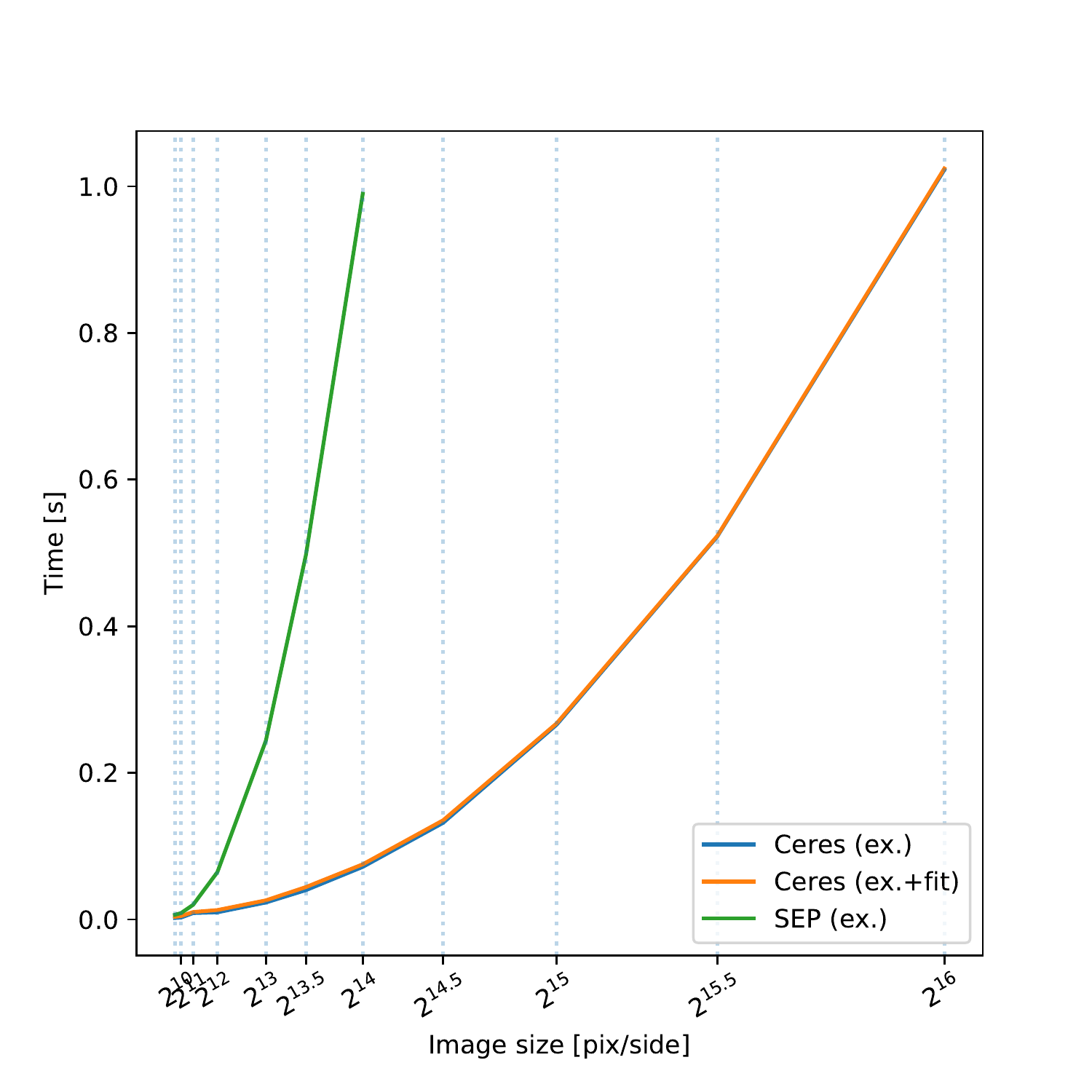}
\caption{Comparative benchmarks for extraction of 64 sources on images of varying size - C/C++ packages only. 
Some experiments perform only source-extraction (ex.), while others perform both source-extraction and fitting (ex.+fit).}
\label{fig:comparative_extraction_c}
\end{figure}

In addition to the detailed benchmarking of our C++ implementation described above, we also performed some comparative benchmarks with the other widely-used source-finding packages, some of which were outlined in \S~\ref{sec:other_sourcefinders}, namely:
\begin{itemize}
 \item PySE - LOFAR-TKP Python Source-Extractor \cite{spreeuw_pyse, swinbank_trap};
 \item SEP - fork of the S-Extractor package \cite{bertin_sextractor};
 \item Python ref. code - reference Python code we implemented as a guide for the C++ implementation;
 \item {\sc Ceres} - the C++ source find algorithm proposed in this paper;
\end{itemize}

Test-data consisted of randomly generated images of different sizes containing 64 gaussian-shaped sources. The generated test images were correctly shifted, so the labelling algorithm was compiled considering this fact, \emph{i.e.} these tests are not accounting for the additional gains that can be achieved by avoiding the quadrant shift operation after FFT usually performed in the radio imaging pipeline.

For these benchmarks we focused on two key stages, the background and noise estimation, and source-extraction including gaussian fitting. We configured the packages to perform these tasks on in-memory arrays of image-data --- this reflects the likely `real-world' usage of a high-throughput source-extraction pipeline, and removes the contaminating (and potentially benchmark-dominating) effects of different input-output data-access strategies. 
However, the original S-Extractor package \cite{bertin_sextractor} provides only a command-line interface, with no library API for direct calls, and processes images in a piecewise manner to reduce memory overheads. 
We instead benchmark the SEP library \cite{barbary_sep}, a recent fork of S-Extractor which modifies the original S-Extractor codebase to run directly on in-memory arrays and provides Python bindings. We do not provide benchmarks for the Aegean package as the interface provides no easy method for supplying image-arrays directly.
In the interests of simplicity the benchmarks are all run via Python scripts, with C/C++ routines accessed via Python bindings. Consequently, the performance of these packages will be slightly reduced compared to an end-to-end C++ pipeline due to the overhead involved in data-passing. \footnote{It is worth noting that the array-ordering can have a significant influence on performance when using Python bindings --- SEP will only accept arrays in the C-style ordering (the numpy default), but the Python bindings developed for the C++ library described herein are significantly more performant (performance improved by around 1.5--2x) when passed numpy arrays in Fortran ordering, as used internally by the Armadillo routines.}

Figures~\ref{fig:comparative_extraction} and~\ref{fig:comparative_extraction_c} depict the results of timing benchmarks for the source-extraction and fitting process, as performed on images of varying size. The experiments not applying source fitting are labelled with the abbreviation ``\emph{ex.}'', while the full process of source-extraction and fitting is labelled with ``\emph{ex.+fit}''.
Figure~\ref{fig:comparative_extraction} shows the results for all packages tested.
The first striking result is how much slower the Python implementations are compared to the C/C++ packages --- this is unexpected given that the Python implementations make heavy use of the Numpy / Scipy libraries which in turn make heavy use of C-extensions. We can observe that the improvement from a dedicated C++ routine is fairly dramatic, with a speedup factor of several hundreds.
We also note that performance of our reference Python implementation is significantly slower than PySE --- this seems to be due to performing additional array copies and highlights the challenges of working efficiently with large arrays in Python. 

Figure~\ref{fig:comparative_extraction_c} depicts the results for the C/C++ packages only, with tests performed at varying image sizes of widths from $2^{10}$ -- $2^{14}$ (1024 -- 16384) pixels. The novel C++ implementation is markedly faster than SEP for large image sizes, completing the source-extraction approximately 13 times faster than SEP for an image width of  $2^{14}$ pixels. 
For the proposed {\sc Ceres} package we plot two lines, one for extraction and moments-estimation alone --- which is directly comparable to the results given by SEP --- and a second line which also includes the time required for Gaussian fitting of each source-island. 
Since the number of sources is constant, we expect an efficient implementation of the island-fitting to perform in constant time for a constant number of sources regardless of the image size, and this is true to a good approximation for the {\sc Ceres} implementation, with the fitting procedure adding a near-constant and quite small additional execution time across image-sizes (some milliseconds). 
We note that we were unable to perform benchmarks of SEP at image-widths above $2^{14}$ as attempting to do so resulted in a segmentation fault.

It is important to note that the timing results of the proposed {\sc Ceres} package in these experiments include the parallel processing techniques. However, further experiments using a single thread revealed that the developed package continues presenting a superior performance than SEP.

\section{Conclusions}
\label{sec:conclusions}

Upcoming observatories, such as the Square Kilometre Array, demand high efficient algorithms for object detection in astronomical images. In this paper, we discuss a new high performant solution for source finding, which incorporates a number of advanced techniques to accelerate its algorithm and adapt it for multi-core processing systems. 

Main contributions include: fast median computation for background noise estimation based on data binning which presents better parallel scalability than traditional quick-select approach; fast sigma clipping algorithm for RMS estimation which minimises the amount of computations and memory accesses; and an efficient and highly parallelisable CCL algorithm that supports detection of positive and negative sources with reduced computational complexity and processes non-shifted images as outputted by FFT from the previous imaging pipeline stage.
Timing benchmarks demonstrate the superior computational performance of the techniques proposed for source finding, in particular when using multi-core processing. 
Comparative benchmarks against existing source finding algorithms show that the developed C++ solution presents a state-of-the-art performance when processing in-memory arrays of image-data.

\section*{Acknowledgements}

The authors gratefully acknowledge support from the UK Science and Technology Facilities Council.

The authors gratefully acknowledge the support of the ENGAGE SKA HPC cluster at IT-Aveiro. ENGAGE SKA (POCI-01-0145-FEDER-022217) is funded by COMPETE 2020 and FCT, Portugal.

The authors gratefully acknowledge the developers of the following packages that were used in this work: Armadillo\footnote{http://arma.sourceforge.net} \cite{Curtin_JOS2016}, Google Test\footnote{https://github.com/google/googletest}, Google Benchmark\footnote{https://github.com/google/benchmark}, cnpy\footnote{https://github.com/rogersce/cnpy}, FFTW3\footnote{http://www.fftw.org} \cite{Frigo_FFTW05}, pybind11\footnote{https://github.com/pybind/pybind11}, TBB\footnote{https://www.threadingbuildingblocks.org} \cite{Reinders_TBB}, RapidJSON\footnote{https://github.com/miloyip/rapidjson}, TCLAP\footnote{http://tclap.sourceforge.net}, spdlog\footnote{https://github.com/gabime/spdlog}, Eigen\footnote{http://eigen.tuxfamily.org} \cite{eigenweb}, Ceres Solver\footnote{http://ceres-solver.org}, Astropy\footnote{http://www.astropy.org} \cite{astropy:2013, astropy:2018}, Numpy\footnote{http://www.numpy.org} \cite{numpy}, Scipy\footnote{http://www.scipy.org} \cite{scipy}, Matplotlib\footnote{https://matplotlib.org/} \cite{Hunter:2007}, Jupyter Notebook\footnote{https://jupyter.org/}, Spack\footnote{https://spack.io} \cite{Gamblin:2015}, GNU/Linux operating system and tools\footnote{https://www.gnu.org}.





\section*{References}
\bibliographystyle{model1-num-names}
\bibliography{bibtex/stp_bibfile}

\begin{thebibliography}{36}
\expandafter\ifx\csname natexlab\endcsname\relax\def\natexlab#1{#1}\fi
\providecommand{\bibinfo}[2]{#2}
\ifx\xfnm\relax \def\xfnm[#1]{\unskip,\space#1}\fi
\bibitem[{Hancock et~al.(2012)Hancock, Murphy, Gaensler, Hopkins, and
  Curran}]{hancock_aegean}
\bibinfo{author}{P.~J. Hancock}, \bibinfo{author}{T.~Murphy},
  \bibinfo{author}{B.~M. Gaensler}, \bibinfo{author}{A.~Hopkins},
  \bibinfo{author}{J.~R. Curran},
\newblock \bibinfo{title}{Compact continuum source finding for next generation
  radio surveys},
\newblock \bibinfo{journal}{Monthly Notices of the Royal Astronomical Society}
  \bibinfo{volume}{422} (\bibinfo{year}{2012}) \bibinfo{pages}{1812--1824}.
\bibitem[{Jonas(2009)}]{MeerKAT}
\bibinfo{author}{J.~L. Jonas},
\newblock \bibinfo{title}{The south african array with composite dishes and
  wide-band single pixel feeds},
\newblock \bibinfo{journal}{Proceedings of the IEEE} \bibinfo{volume}{97}
  (\bibinfo{year}{2009}) \bibinfo{pages}{1522--1530}.
\bibitem[{et~al.(2009)}]{ASKAP}
\bibinfo{author}{D.~R.~D. et~al.},
\newblock \bibinfo{title}{Australian ska pathfinder: A high-dynamic range
  wide-field of view survey telescope},
\newblock \bibinfo{journal}{Proceedings of the IEEE} \bibinfo{volume}{97}
  (\bibinfo{year}{2009}) \bibinfo{pages}{1507--1521}.
\bibitem[{Dewdney et~al.(2009)Dewdney, Hall, Schilizzi, and Lazio}]{SKA}
\bibinfo{author}{P.~Dewdney}, \bibinfo{author}{P.~J. Hall},
  \bibinfo{author}{R.~T. Schilizzi}, \bibinfo{author}{T.~J. L.~W. Lazio},
\newblock \bibinfo{title}{The square kilometre array},
\newblock \bibinfo{journal}{Proceedings of the IEEE} \bibinfo{volume}{97}
  (\bibinfo{year}{2009}) \bibinfo{pages}{1482--1496}.
\bibitem[{Software(2018)}]{STP_website}
\bibinfo{author}{C.~Software}, \bibinfo{title}{Slow transients pipeline
  prototype - c++ implementation}, \bibinfo{year}{2018}.
\bibitem[{Cornwell(2008)}]{CLEAN}
\bibinfo{author}{T.~Cornwell},
\newblock \bibinfo{title}{Multiscale clean deconvolution of radio synthesis
  images},
\newblock \bibinfo{journal}{IEEE Journal of Selected Topics in Signal
  Processing} \bibinfo{volume}{2} (\bibinfo{year}{2008})
  \bibinfo{pages}{793--801}.
\bibitem[{{Mart{\'{\i}}-Vidal} et~al.(2014){Mart{\'{\i}}-Vidal}, {Vlemmings},
  {Muller}, and {Casey}}]{martividal_uvmultifit}
\bibinfo{author}{I.~{Mart{\'{\i}}-Vidal}}, \bibinfo{author}{W.~H.~T.
  {Vlemmings}}, \bibinfo{author}{S.~{Muller}}, \bibinfo{author}{S.~{Casey}},
\newblock \bibinfo{title}{{UVMULTIFIT: A versatile tool for fitting
  astronomical radio interferometric data}},
\newblock \bibinfo{journal}{\aap} \bibinfo{volume}{563} (\bibinfo{year}{2014})
  \bibinfo{pages}{A136}.
\bibitem[{{Lochner} et~al.(2015){Lochner}, {Natarajan}, {Zwart}, {Smirnov},
  {Bassett}, {Oozeer}, and {Kunz}}]{lochner_biro}
\bibinfo{author}{M.~{Lochner}}, \bibinfo{author}{I.~{Natarajan}},
  \bibinfo{author}{J.~T.~L. {Zwart}}, \bibinfo{author}{O.~{Smirnov}},
  \bibinfo{author}{B.~A. {Bassett}}, \bibinfo{author}{N.~{Oozeer}},
  \bibinfo{author}{M.~{Kunz}},
\newblock \bibinfo{title}{{Bayesian inference for radio observations}},
\newblock \bibinfo{journal}{\mnras} \bibinfo{volume}{450}
  (\bibinfo{year}{2015}) \bibinfo{pages}{1308--1319}.
\bibitem[{{Bertin} and {Arnouts}(1996)}]{bertin_sextractor}
\bibinfo{author}{E.~{Bertin}}, \bibinfo{author}{S.~{Arnouts}},
\newblock \bibinfo{title}{{SExtractor: Software for source extraction.}},
\newblock \bibinfo{journal}{\aaps} \bibinfo{volume}{117} (\bibinfo{year}{1996})
  \bibinfo{pages}{393--404}.
\bibitem[{{Spreeuw}(2010)}]{spreeuw_pyse}
\bibinfo{author}{J.~N. {Spreeuw}}, \bibinfo{title}{{Search and detection of low
  frequency radio transients}}, Ph.D. thesis, {University of Amsterdam},
  \bibinfo{year}{2010}.
\bibitem[{{Swinbank} et~al.(2015){Swinbank}, {Staley}, {Molenaar}, {Rol},
  {Rowlinson}, {Scheers}, {Spreeuw}, {Bell}, {Broderick}, {Carbone}, {Garsden},
  {van der Horst}, {Law}, {Wise}, {Breton}, {Cendes}, {Corbel},
  {Eisl{\"o}ffel}, {Falcke}, {Fender}, {Grie{\ss}meier}, {Hessels}, {Stappers},
  {Stewart}, {Wijers}, {Wijnands}, and {Zarka}}]{swinbank_trap}
\bibinfo{author}{J.~D. {Swinbank}}, \bibinfo{author}{T.~D. {Staley}},
  \bibinfo{author}{G.~J. {Molenaar}}, \bibinfo{author}{E.~{Rol}},
  \bibinfo{author}{A.~{Rowlinson}}, \bibinfo{author}{B.~{Scheers}},
  \bibinfo{author}{H.~{Spreeuw}}, \bibinfo{author}{M.~E. {Bell}},
  \bibinfo{author}{J.~W. {Broderick}}, \bibinfo{author}{D.~{Carbone}},
  \bibinfo{author}{H.~{Garsden}}, \bibinfo{author}{A.~J. {van der Horst}},
  \bibinfo{author}{C.~J. {Law}}, \bibinfo{author}{M.~{Wise}},
  \bibinfo{author}{R.~P. {Breton}}, \bibinfo{author}{Y.~{Cendes}},
  \bibinfo{author}{S.~{Corbel}}, \bibinfo{author}{J.~{Eisl{\"o}ffel}},
  \bibinfo{author}{H.~{Falcke}}, \bibinfo{author}{R.~{Fender}},
  \bibinfo{author}{J.-M. {Grie{\ss}meier}}, \bibinfo{author}{J.~W.~T.
  {Hessels}}, \bibinfo{author}{B.~W. {Stappers}}, \bibinfo{author}{A.~J.
  {Stewart}}, \bibinfo{author}{R.~A.~M.~J. {Wijers}},
  \bibinfo{author}{R.~{Wijnands}}, \bibinfo{author}{P.~{Zarka}},
\newblock \bibinfo{title}{{The LOFAR Transients Pipeline}},
\newblock \bibinfo{journal}{Astronomy and Computing} \bibinfo{volume}{11}
  (\bibinfo{year}{2015}) \bibinfo{pages}{25--48}.
\bibitem[{Hancock et~al.(2018)Hancock, Trott, and
  Hurley-Walker}]{hancock_aegean_2018}
\bibinfo{author}{P.~J. Hancock}, \bibinfo{author}{C.~M. Trott},
  \bibinfo{author}{N.~Hurley-Walker},
\newblock \bibinfo{title}{Source finding in the era of the ska (precursors):
  Aegean 2.0},
\newblock \bibinfo{journal}{Publications of the Astronomical Society of
  Australia} \bibinfo{volume}{35} (\bibinfo{year}{2018}) \bibinfo{pages}{e011}.
\bibitem[{Benjamini and Hochberg(1995)}]{benjamini_fdr}
\bibinfo{author}{Y.~Benjamini}, \bibinfo{author}{Y.~Hochberg},
\newblock \bibinfo{title}{Controlling the false discovery rate: A practical and
  powerful approach to multiple testing},
\newblock \bibinfo{journal}{Journal of the Royal Statistical Society. Series B
  (Methodological)} \bibinfo{volume}{57} (\bibinfo{year}{1995})
  \bibinfo{pages}{289--300}.
\bibitem[{{Hopkins} et~al.(2002){Hopkins}, {Miller}, {Connolly}, {Genovese},
  {Nichol}, and {Wasserman}}]{hopkins_fdr}
\bibinfo{author}{A.~M. {Hopkins}}, \bibinfo{author}{C.~J. {Miller}},
  \bibinfo{author}{A.~J. {Connolly}}, \bibinfo{author}{C.~{Genovese}},
  \bibinfo{author}{R.~C. {Nichol}}, \bibinfo{author}{L.~{Wasserman}},
\newblock \bibinfo{title}{{A New Source Detection Algorithm Using the
  False-Discovery Rate}},
\newblock \bibinfo{journal}{\aj} \bibinfo{volume}{123} (\bibinfo{year}{2002})
  \bibinfo{pages}{1086--1094}.
\bibitem[{Lutz(1980)}]{lutz_onepass_labelling}
\bibinfo{author}{R.~K. Lutz},
\newblock \bibinfo{title}{An algorithm for the real time analysis of digitised
  images},
\newblock \bibinfo{journal}{The Computer Journal} \bibinfo{volume}{23}
  (\bibinfo{year}{1980}) \bibinfo{pages}{262--269}.
\bibitem[{Jones et~al.(01  )Jones, Oliphant, Peterson et~al.}]{scipy}
\bibinfo{author}{E.~Jones}, \bibinfo{author}{T.~Oliphant},
  \bibinfo{author}{P.~Peterson}, et~al., \bibinfo{title}{{SciPy}: Open source
  scientific tools for {Python}}, \bibinfo{year}{2001--}.
\bibitem[{Floyd and Rivest(1975)}]{Floyd_quickselect}
\bibinfo{author}{R.~W. Floyd}, \bibinfo{author}{R.~L. Rivest},
\newblock \bibinfo{title}{Algorithm 489: The algorithm select - for finding the
  ith smallest of n elements [m1]},
\newblock \bibinfo{journal}{Commun. ACM} \bibinfo{volume}{18}
  (\bibinfo{year}{1975}).
\bibitem[{Singler and Konsik(2008)}]{gnu_std_parallel}
\bibinfo{author}{J.~Singler}, \bibinfo{author}{B.~Konsik},
\newblock \bibinfo{title}{The gnu libstdc++ parallel mode: Software engineering
  considerations},
\newblock in: \bibinfo{booktitle}{Proceedings of the 1st International Workshop
  on Multicore Software Engineering}, pp. \bibinfo{pages}{15--22}.
\bibitem[{Tibshirani(2008)}]{Tibshirani_binmedian}
\bibinfo{author}{R.~J. Tibshirani},
\newblock \bibinfo{title}{Fast computation of the median by successive
  binning},
\newblock \bibinfo{journal}{CoRR}  (\bibinfo{year}{2008}).
\bibitem[{Reinders(2007)}]{Reinders_TBB}
\bibinfo{author}{J.~Reinders}, \bibinfo{title}{Intel Threading Building
  Blocks}, \bibinfo{publisher}{O'Reilly \& Associates, Inc.},
  \bibinfo{address}{Sebastopol, CA, USA}, \bibinfo{edition}{first} edition,
  \bibinfo{year}{2007}.
\bibitem[{Samet and Tamminen(1988)}]{Samet_CCL_1988}
\bibinfo{author}{H.~Samet}, \bibinfo{author}{M.~Tamminen},
\newblock \bibinfo{title}{Efficient component labeling of images of arbitrary
  dimension represented by linear bintrees},
\newblock \bibinfo{journal}{IEEE Transactions on Pattern Analysis and Machine
  Intelligence} \bibinfo{volume}{10} (\bibinfo{year}{1988})
  \bibinfo{pages}{579--586}.
\bibitem[{Jung and Jeong(2010)}]{Jeong_ISCIT2010}
\bibinfo{author}{I.~Y. Jung}, \bibinfo{author}{C.~S. Jeong},
\newblock \bibinfo{title}{Parallel connected-component labeling algorithm for
  gpgpu applications},
\newblock in: \bibinfo{booktitle}{2010 10th International Symposium on
  Communications and Information Technologies}, pp.
  \bibinfo{pages}{1149--1153}.
\bibitem[{Cabaret et~al.(2016)Cabaret, Lacassagne, and Etiemble}]{Cabaret2016}
\bibinfo{author}{L.~Cabaret}, \bibinfo{author}{L.~Lacassagne},
  \bibinfo{author}{D.~Etiemble},
\newblock \bibinfo{title}{Parallel light speed labeling: an efficient connected
  component algorithm for labeling and analysis on multi-core processors},
\newblock \bibinfo{journal}{Journal of Real-Time Image Processing}
  (\bibinfo{year}{2016}) \bibinfo{pages}{1--24}.
\bibitem[{Grana et~al.(2018)Grana, Bolelli, Baraldi, and
  Vezzani}]{yacclab_website}
\bibinfo{author}{C.~Grana}, \bibinfo{author}{F.~Bolelli},
  \bibinfo{author}{L.~Baraldi}, \bibinfo{author}{R.~Vezzani},
  \bibinfo{title}{Yacclab: Yet another connected components labeling
  benchmark}, \bibinfo{howpublished}{https://github.com/prittt/YACCLAB},
  \bibinfo{year}{2018}.
\bibitem[{Bradski(2000)}]{opencv_library}
\bibinfo{author}{G.~Bradski},
\newblock \bibinfo{title}{{The OpenCV Library}},
\newblock \bibinfo{journal}{Dr. Dobb's Journal of Software Tools}
  (\bibinfo{year}{2000}).
\bibitem[{Wu et~al.(2008)Wu, Otoo, and Suzuki}]{Wu_CCL_openCV}
\bibinfo{author}{K.~Wu}, \bibinfo{author}{E.~Otoo},
  \bibinfo{author}{K.~Suzuki},
\newblock \bibinfo{title}{Two strategies to speed up connected component
  labeling algorithms},
\newblock \bibinfo{journal}{Lawrence Berkeley National Laboratory}
  (\bibinfo{year}{2008}).
\bibitem[{Agarwal et~al.(2010)Agarwal, Mierle et~al.}]{ceressolver}
\bibinfo{author}{S.~Agarwal}, \bibinfo{author}{K.~Mierle}, et~al.,
  \bibinfo{title}{Ceres solver},
  \bibinfo{howpublished}{http://ceres-solver.org}, \bibinfo{year}{2010}.
\bibitem[{Sanderson and Curtin(2016)}]{Curtin_JOS2016}
\bibinfo{author}{C.~Sanderson}, \bibinfo{author}{R.~Curtin},
\newblock \bibinfo{title}{Armadillo: a template-based c++ library for linear
  algebra},
\newblock \bibinfo{journal}{Journal of Open Source Software}
  \bibinfo{volume}{1} (\bibinfo{year}{2016}) \bibinfo{pages}{26}.
\bibitem[{Barbary(2016)}]{barbary_sep}
\bibinfo{author}{K.~Barbary},
\newblock \bibinfo{title}{{SEP}: Source extractor as a library},
\newblock \bibinfo{journal}{The Journal of Open Source Software}
  \bibinfo{volume}{1} (\bibinfo{year}{2016}).
\bibitem[{Frigo and Johnson(2005)}]{Frigo_FFTW05}
\bibinfo{author}{M.~Frigo}, \bibinfo{author}{S.~G. Johnson},
\newblock \bibinfo{title}{The design and implementation of {FFTW3}},
\newblock \bibinfo{journal}{Proceedings of the IEEE} \bibinfo{volume}{93}
  (\bibinfo{year}{2005}) \bibinfo{pages}{216--231}.
\bibitem[{Guennebaud et~al.(2010)Guennebaud, Jacob et~al.}]{eigenweb}
\bibinfo{author}{G.~Guennebaud}, \bibinfo{author}{B.~Jacob}, et~al.,
  \bibinfo{title}{Eigen v3},
  \bibinfo{howpublished}{http://eigen.tuxfamily.org}, \bibinfo{year}{2010}.
\bibitem[{{Astropy Collaboration} et~al.(2013){Astropy Collaboration},
  {Robitaille} et~al.}]{astropy:2013}
\bibinfo{author}{{Astropy Collaboration}}, \bibinfo{author}{T.~P.
  {Robitaille}}, et~al.,
\newblock \bibinfo{title}{{Astropy: A community Python package for astronomy}},
\newblock \bibinfo{journal}{aap} \bibinfo{volume}{558} (\bibinfo{year}{2013})
  \bibinfo{pages}{A33}.
\bibitem[{{Price-Whelan} et~al.(2018)}]{astropy:2018}
\bibinfo{author}{A.~M. {Price-Whelan}}, et~al.,
\newblock \bibinfo{title}{{The Astropy Project: Building an Open-science
  Project and Status of the v2.0 Core Package}},
\newblock \bibinfo{journal}{aj} \bibinfo{volume}{156} (\bibinfo{year}{2018})
  \bibinfo{pages}{123}.
\bibitem[{Oliphant(06  )}]{numpy}
\bibinfo{author}{T.~Oliphant}, \bibinfo{title}{{NumPy}: A guide to {NumPy}},
  \bibinfo{howpublished}{USA: Trelgol Publishing}, \bibinfo{year}{2006--}.
\bibitem[{Hunter(2007)}]{Hunter:2007}
\bibinfo{author}{J.~D. Hunter},
\newblock \bibinfo{title}{Matplotlib: A 2d graphics environment},
\newblock \bibinfo{journal}{Computing In Science \& Engineering}
  \bibinfo{volume}{9} (\bibinfo{year}{2007}) \bibinfo{pages}{90--95}.
\bibitem[{Gamblin et~al.(2015)Gamblin, LeGendre, Collette, Lee, Moody,
  de~Supinski, and Futral}]{Gamblin:2015}
\bibinfo{author}{T.~Gamblin}, \bibinfo{author}{M.~LeGendre},
  \bibinfo{author}{M.~R. Collette}, \bibinfo{author}{G.~L. Lee},
  \bibinfo{author}{A.~Moody}, \bibinfo{author}{B.~R. de~Supinski},
  \bibinfo{author}{S.~Futral},
\newblock \bibinfo{title}{The spack package manager: Bringing order to {HPC}
  software chaos},
\newblock \bibinfo{journal}{Proceedings of the International Conference for
  High Performance Computing, Networking, Storage and Analysis}
  (\bibinfo{year}{2015}) \bibinfo{pages}{40:1--40:12}.

\end{thebibliography}







\end{document}